\documentclass[12pt]{article}

\usepackage{amssymb}
\usepackage{amsmath}
\usepackage[hyphens]{url}
\usepackage[title,toc,page]{appendix}
\usepackage{graphicx}
\usepackage{subcaption}
\usepackage[margin=1in]{geometry}
\usepackage{multicol}
\usepackage{siunitx}
\usepackage{mathrsfs}

\usepackage{booktabs}  
\usepackage{float} 
\usepackage{tabularx,ragged2e}
\usepackage[tablename=Table,aboveskip=5pt]{caption}
\usepackage{relsize}
\usepackage[space]{cite}

\begin{document}

\begin{titlepage}

\begin{center}

\begin{flushright}
\end{flushright}
\vskip 2.5cm

{\Large \bf Exiting Inflation with a Smooth Scale Factor}
\end{center}

\vspace{1ex}

\begin{center}
{\large Harry Oslislo and Brett Altschul\footnote{{\tt altschul@mailbox.sc.edu}}}

\vspace{5mm}
{\sl Department of Physics and Astronomy} \\
{\sl University of South Carolina} \\
{\sl Columbia, SC 29208} \\
\end{center}

\vspace{2.5ex}

\medskip

\centerline {\bf Abstract}

\bigskip

The expectation that the physical expansion of space occurs smoothly
may be expressed mathematically as a requirement for continuity in the time derivative of the
metric scale factor of the Friedmann–Robertson–Walker cosmology.
We explore the consequences of imposing such a smoothness requirement, examining the forms of possible
interpolating functions between the end of inflation and subsequent radiation- or matter-dominated eras,
using a straightforward geometric model of the interpolating behavior. We quantify the magnitude of
the cusp found in a direct transition from the end of slow roll inflation to the subsequent era,
analyze the validity several smooth interpolator candidates, and investigate equation-of-state and
thermodynamic constraints. We find an order-of-magnitude increase in the size of the universe at
the end of the transition to a single-component radiation or matter era. We also evaluate the
interpolating functions in terms of the standard theory of preheating and determine the effect on
the number of bosons produced.

\bigskip

\end{titlepage}

\newpage

\section{Introduction}

The problem of
trying to reconcile physical theories regarding the form and evolution of the primordial universe with
modern cosmological observations has occupied researchers for decades. Providing a complete explanation
for the origins of the characteristics that the universe exhibits today has been challenging. We witness
extreme uniformity and flatness and the absence of certain particles predicted in some Grand Unified Theory
(GUT) models. Many physicists have detailed these well-known difficulties in texts and expository
papers; see, for example, Refs.~\cite{LiddleLyth2006,ryden2003,linde1984,baumann2012,lesgourgues2006,
liddle2015,bassett2006,kinney2004,weinberg2008}.

The issue of uniformity has the name the \textit{Horizon Problem}. We see a homogeneous,
isotropic universe on large scales. Despite the almost incomprehensible longevity of the universe,
it is simply too immense to have grown to be uniform on large scales. Causality demands that local
homogeneity takes time to develop, with equilibrium conditions dispersing at a rate no greater than the
speed of light. Two local homogeneous elements dispersed in time must retain a causal connection to
remain in equilibrium with one another. Yet opposite sides of the universe appear nearly identical to us.
Even if we assume they started that way, not enough time has passed for space to have expanded a
great enough distance to maintain the equilibrium---at least not according to the physics we
understand---with signaling bounded by the limit of the speed of light. Cosmologists assess uniformity
primarily using the temperature of the Cosmic Microwave Background (CMB), which is the thermal radiation
emitted as the matter in the universe was cooling and transitioning from a conductive, opaque plasma to a
neutral, transparent gas. The widely-accepted standard is $T_{\text{CMB}} \approx 2.7255$K~\cite{fixsen2009}.
An early CMB probe, the Cosmic Background Explorer, found the temperature variation from this mean to be on
the order of $10^{-5}$ K. Each causally connected portion of the CMB that should have been able
to thermalize before the recombination photons began to stream freely along their paths toward the Earth
covers a solid angle of approximately $0.013$ sr on the sky, so that about $10^{4}$ such solid angles
make up the CMB. \textit{How then,} cosmologists ask, \textit{did ten thousand discrete portions of the CMB,
which do not appear to have been causally connected at the time of recombination, collectively equilibrate at
a common temperature that is uniform to within the order of $10^{-5}$ K}~\cite{ryden2003}? A period
of superluminal expansion provided by inflation could provide the missing causal connection.

Expansion itself does not change the inherent topology of the universe: A universe that is
closed, open, or flat remains so. However, expansion makes any curvature appear locally more flat.
The Planck Collaboration has measured the spatial curvature $\Omega_k$ as $0.001\pm 0.002$~\cite{PlIV2018},
which means our nearly flat universe presents a second cosmological difficulty, the \textit{Flatness Problem}.
The scale factors $a(t)$ of matter-dominated and radiation-dominated cosmologies are proportional to $t^{2/3}$
and $t^{1/2}$, respectively. According to the first Friedman equation~\cite{friedmann1922, ryden2003},
$\Omega _k$ scales as $1/\dot{a}^2$, so that in the absence of any other influences, the longevity of the
universe means consistency with the Planck measurement requires that the curvature at the beginning of
the radiation-dominated era must have been extraordinarily small. Again, inflation offers a remedy:
The inflationary exponential scale factor would tend to drive down the spatial curvature to a level that
could support subsequent evolution to the value observed today.

The additional {\em Monopole Problem} arises from the predictions of some GUT
models~\cite{hooft1974, polyakov1974} that a phase transition breaks the symmetry between the strong
and weak forces when the temperature of the universe drops to a level consistent with the energy scale
$10^{16}$ GeV. A result would be the formation of a dust of massive magnetic monopoles, with a density that is
subsequently proportional to $a^{-3}$, potentially thereby blocking the radiation and matter eras from
taking place~\cite{lesgourgues2006}. The Monopole Problem calls for a mechanism to reconcile the GUT
prediction of the creation of these massive particles with our accepted understanding of the chronology
of the early universe and current cosmological observation. Inflation could provide dilution that would
make magnetic monopoles so few and far apart that finding them would be essentially impossible.

\subsection{The Inflation Solution}

In his groundbreaking paper in 1981~\cite{guth1981}, Alan Guth introduced inflation as a theory to
address the inexplicable Horizon, Flatness, and Monopole Problems. However, he also acknowledged the
difficulty his mechanism created: An exit from the false vacuum that drives inflation involved quantum
tunneling from a false to the true vacuum state, an effect that would occur primarily in localized
bubbles---that is, discrete regions subsequently characterized by the Klein-Gordon scalar field that
drives inflation (the inflaton $\phi$) having settled
into its true vacuum state. Meanwhile, expansion of space would continue between the bubbles (where such
tunneling had not yet occurred), and as a result, we would expect to see parcels of non-uniform space today.
Intersecting bubbles would have similar effects. This model of inflation thus predicted a universe
inconsistent with observation; Guth's original theory lacks a \textit{graceful exit}.

\begin{figure}[ht!]
    \centering
    \subfloat[\centering]{{\includegraphics[width=7cm]{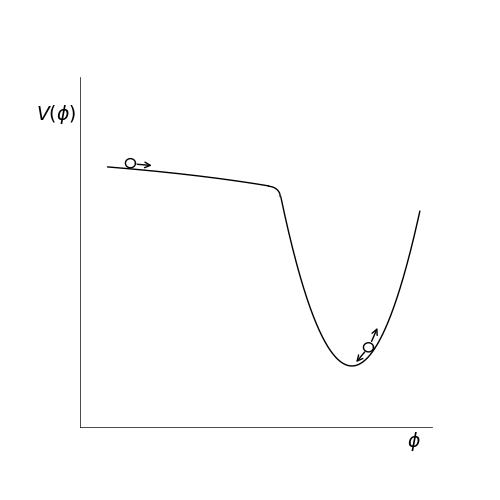} }}%
    \qquad
    \subfloat[\centering]{{\includegraphics[width=7cm]{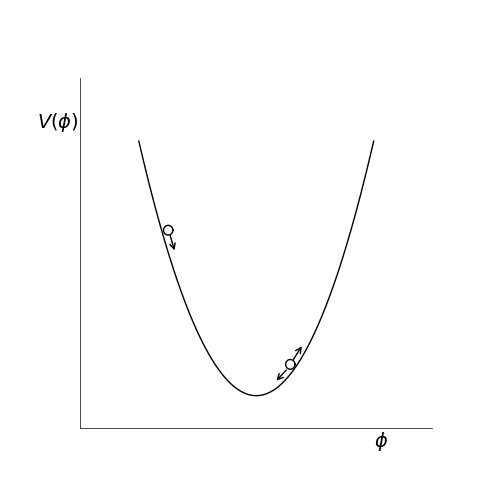} }}%
		\caption{The contrast between the potentials of slow roll inflation, shown in graph (a), and
		chaotic inflation, in graph (b). Slow roll inflation requires a plateau to generate enough
		$e$-folds of inflationary expansion to solve the Horizon, Flatness, and Monopole Problems.
		In chaotic inflation, the effects of the friction term in the equation of motion replace that
		of the plateau in keeping the inflaton from moving to the true vacuum too quickly. After inflation,
		both models involve the inflaton oscillating around a minimum potential during a period of
		reheating, which we review in section~\ref{Reheating}.}%
    \label{potbypot}
\end{figure}

In 1982, inflation pioneer Andrei Linde sought to solve the graceful exit problem with a new theory,
\textit{slow-roll inflation}~\cite{linde1982}. Instead of starting in a false vacuum, the inflaton rolls
down a potential energy plateau to a minimum where it oscillates around a true vacuum state,
which graph (a) of figure~\ref{potbypot} depicts schematically. The assumption that the potential
energy of the inflaton dominates the kinetic energy for a sufficient time results in exponential inflation.

The \textit{slow-roll} condition implies $\ddot{\phi} \rightarrow 0$, so that the equation of motion
of the scalar field,
\begin{equation}\label{eomft}
\ddot \phi + 3H\dot \phi +V(\phi)_{,\phi} = 0, 
\end{equation}
reduces to
\begin{equation}
3H\dot \phi= -V(\phi)_{,\phi}.
\end{equation}
The Hubble parameter $H$ is the expansion rate of the universe,
$V(\phi)$ is the potential of the inflaton, and the comma denotes the partial derivative with respect
to $\phi$. The theory assumes that the magnitudes of both the density and pressure of the inflaton
become approximately equal to the potential by treating the inflaton condensate as a perfect fluid:
\begin{align}
\rho &= \frac{1}{2} \dot \phi^2 + V(\phi) \approx V(\phi) \\
p &= \frac{1}{2} \dot \phi^2 - V(\phi) \approx -V(\phi).
\end{align}
In this regime, the second Friedmann equation~\cite{friedmann1922,ryden2003}, commonly known as the
\textit{acceleration equation}, because it governs $\ddot{a}$, is essentially
\begin{equation}\label{acceq}
\frac{\ddot{a}(t)}{a(t)} = -\frac{4\pi G}{3}(\rho + 3p) = \frac{8\pi G}{3}\rho.
\end{equation}
In our notation, $\hbar = 1$ and $c = 1$ hereafter unless otherwise noted. Thus the expansion of space
undergoes inflationary acceleration, $\frac{\ddot a}{a} > 0$, as a result. The first Friedman equation,
\begin{equation}
H^2 + \frac{k}{a^2}= \frac{8\pi G}{3}\rho,
\end{equation}
with $H^2 = \big( \frac{\dot a}{a} \big)^2 \gg \frac{k}{a^2}$, yields the scale factor solution 
\begin{equation}\label{ainfl}
a(t) = a_0e^{Ht}.
\end{equation}
This is the exponential expansion of space predicted by the theory of slow-roll
inflation~\cite{kinney2004, linde2007}. A period of superluminal expansion would explain the homogeneity
and isotropy of the observable universe by providing the necessary causal connection to solve the
Horizon Problem. Superluminal expansion would also flatten the spatial curvature and decrease the
density of magnetic monopoles. Numerical analysis provides insight into the question of the number of
$e$-folds of expansion necessary to resolve these problems. However, this solution comes with its own
associated shortcoming: Producing an outcome consistent with modern observations demands very specific
initial conditions.

Although the underlying physics of inflation (such as the existence of the inflaton field) remains
unsubstantiated experimentally, the framework of inflation is widely accepted among cosmologists as a way
of providing an underlying solution to various cosmological problems. Over the years, researchers have
revised the concept by devising a diverse body of new theories. For our purposes, we shall focus on
Linde's solution to the problem of the requirement of specific initial conditions in the slow roll theory.
In 1983, he published his theory of \textit{chaotic inflation}~\cite{linde1983}. In its simplest
version~\cite{linde2007}, the inflaton potential has the form $V(\phi) = \frac{1}{2}m^2 \phi^2$.
The plateau is absent, and in an expanding universe, the friction term $3H\dot \phi$ in the inflaton
equation of motion (\ref{eomft}) has the effect of restricting the motion of the inflaton, as the slow
roll plateau does, resulting again in exponential expansion. Figure~\ref{potbypot} shows the two
contrasting potentials.

Cosmologists have applied a variety of approaches to estimating the amount of inflation necessary to
solve the Horizon, Flatness, and Monopole Problems. The amount by which the cosmos expands is normally
expressed in terms of the number of times the size has increased by a constant factor---in
other words, the number of the $e$-folds (or nepers) $N = \log\frac{a(t_f)}{a(t_i)}$.
Linde~\cite{linde2007} reports
that a quadratic inflaton potential creates a wavelength for the inflaton comparable in size to our
observable universe after about 61 $e$-folds. In a detailed analysis, Lyth~\cite{lyth2007} finds that for a
quartic inflaton potential, a range of $e$-fold values from 47 to 61 results in a universe on the present
scale. He further explains that at a minimum, more than 14 $e$-folds are needed to generate perturbations
leading to structure formation, and that an extended period of domination of the inflaton kinetic term could
increase his estimates of $N$ by nearly 14, therefore concluding with an estimate of $14 < N < 75$.

Other researchers have performed analyses to determine the number of $e$-folds required to solve
specific inflationary problems~\cite{riotto2017}. Solving the Horizon Problem entails that the comoving
Hubble radius at the beginning of inflation $\left[a(t_i) H(t_i)\right]^{-1}$ must contain what has become the comoving
Hubble radius today $\left[a(t_0) H(t_0)\right]^{-1}$, so that the comoving $\left[a(t_0) H(t_0)\right]^{-1}$ could have
thermalized before expanding through the post-inflationary epochs of the universe up to the present. The
Hubble radius is the distance light travels in time $t = H^{-1}$. Thus we have
\begin{align}
\left[a(t_i) H(t_i)\right]^{-1} & \ge \left[a(t_0) H(t_0)\right]^{-1} \\
\frac{1}{H(t_i)} & \ge \frac{1}{H(t_0)}\frac{a(t_i)}{a(t_0)} = \frac{1}{H(t_0)}\frac{a(t_f)}{a(t_0)}
\frac{a(t_i)}{a(t_f)} = \frac{1}{H(t_0)}\frac{a(t_f)}{a(t_0)} e^{-N} \\
\label{invaT}
-N & \le \log \left[ \frac{H(t_0)}{H(t_i)}\frac{a(t_0)}{a(t_f)} \right] = \log \left[ \frac{H(t_0)}{H(t_i)}
\frac{T(t_f)}{T(t_0)} \right] = \log \left[ \frac{H(t_0)}{T(t_0)}\frac{T(t_f)}{H(t_i)} \right].
\end{align}
In eq.~(\ref{invaT}) we use the inverse relation between the scale factor and temperature, which is derived
in appendix~\ref{temp} for reference. Parameter values $T(t_0) \approx 2.75 \, \text{K}$ and
$H(t_0) \approx 100\,\text{km}/\text{s}/\text{Mpc}$ lead to
\begin{equation}
N \ge 67 + \log \left[ \frac{H(t_i)}{T(t_f)} \right],
\end{equation}
which indicates that $N$ is at least 67, because the temperature $H(t_i)$ represents is greater than $T(t_f)$.

\subsection{Reheating}
\label{Reheating}

The expansion of space by inflation dilutes the number densities of all particles and leaves the universe
cold, with energy concentrated primarily in the inflaton. Following the end of inflation, reheating results
in the transfer of energy from the inflaton to Standard Model particles or their precursors. Reheating has
two stages, first the transfer of energy and then subsequent thermalization to a temperature sufficient to
promote nucleosynthesis of light elements. A mechanism developed by Lev Kofman, Andrei Linde, and Alexei
Starobinsky in their iconic 1997 paper, ``Towards the Theory of Reheating after Inflation''~\cite{kls1997},
which they call \textit{preheating} because it precedes thermalization, supersedes earlier explanations of
reheating by way of perturbation theory and narrow parametric resonance. We reference a selection of the
wide range of literature available on the subject of
preheating~\cite{bassett2006,kolb1999,allahverdi2010,linde2002,lozanov2019}.

A preheating framework appears to be necessary, because perturbative processes prove too slow and
inefficient to raise the reheating temperature enough to support nucleosynthesis. Also, the perturbative
approach required certain conditions and treated inflatons collectively in a state of superposition of
individual particles, each capable of decaying independently---rather than as coherent semiclassical fields.
On the other hand, narrow parametric resonance models followed the approach that the inflatons formed a
homogeneous, coherent, oscillating wave appropriate for classical treatment.  In narrow parametric resonance,
an inflaton wave interacts as a background source for a second scalar field $\chi$. However, this theory
itself can be problematic. Because the modes of the scalar field $\chi$ have physical wavelengths, the
expansion of space redshifts modes outside the borders of the resonance band and also makes the band
more narrow. In addition, the expansion and the decay of the inflaton into $\chi$ particles decrease the
amplitude of the coherent inflaton wave. The number of particles being produced instantaneously is
proportional to both the number of $\chi$ particles previously created and to the inflaton amplitude, so
that the effects of expansion and decay lower the efficiency of the resonant conversion and tend to suppress
the growth of the $\chi$ population. Narrow parametric resonance thus typically terminates well before
reheating is complete. 

The parametric resonance in preheating models is instead broad: All modes less than a specific momentum
participate in the $\phi$-$\chi$ coupling. A non-adiabatic transfer of energy leads to exponential growth
in the number and number density of the $\chi$ quanta. Moreover, the expansion of space can actually make the
resonance more effective by gradually redshifting additional modes down to below the maximum momentum,
making them part of the process. The end of reheating depends on the possible range of values of parameters
involved in preheating and the complex dynamics of backreaction and rescattering. However, preheating may
still not be sufficient to complete reheating, and the reheating process may have to revert to a period
of narrow parametric resonance, perturbative decay, or both to arrive at a temperature that is suitable for
thermalization but not high enough to produce very massive particles like monopoles.

In section~\ref{Period Scale Factors}, the reader will find a description of the cusp discontinuity
inherent in inflationary theories involving an exponential scale factor and our approach to quantifying
the extent of the cusp. Section~\ref{The Transition} introduces a method for finding an interpolating
function to replace the cusp, by detailing the geometry of a simple circular model. Then in
section~\ref{Power Law Transitions}, we focus on finding a more realistic interpolating function. We derive
the formalism establishing smoothness in the expansion of space at the end of inflation and analyze the
implications of the most straightforward interpolating candidates, power law functions. The
equation-of-state and thermodynamic constraints provide additional means of restricting possible interpolating
functions, and this is discussed in section~\ref{Additional Constraints}. We analyze the effect on the
size of the universe of a horizontal parabola-like power law serving as a transitional interpolating function
in section~\ref{Results of the Numerical Analysis}. Finally, we look at further numerical analyses to
determine the effect on the scalar $\chi$ number and number density predicted by the Kofman, Linde, and
Starobinsky (KLS) model of preheating in section~\ref{Effect of the Continuous Scale Factor}.

\section{Period Scale Factors}
\label{Period Scale Factors}

The well-known expressions for the scale factor in the early universe include a curious unphysical
approximation, a lack of smoothness at the end of the inflationary epoch. The inflationary and
radiation-dominated scale factors, $a_1 (t)$ and $a_2 (t)$, respectively, follow~\cite{ryden2003}
\begin{align}
a_1 (t) &= a(t_i) e^{H(t-t_i)} \\
a_2 (t) &= a(t_f) \bigg( \frac{t}{t_f} \bigg)^{1/2}. \label{radera}
\end{align} 
To demonstrate the discontinuity in $\dot{a}$, we assume contrariwise that the time derivatives of the
scale factors are equal at the end of inflation, $t_f$:
\begin{align}
\frac{d}{dt}a_1 (t) \,\big|_{t_f} &= a(t_i) H e^{H(t-t_i)} \,\big|_{t_f} = a(t_f) H 
\label{eq-a1der} \\
\frac{d}{dt}a_2 (t) \,\big|_{t_f} &= \frac{a(t_f)}{\sqrt{t_f}} \frac{1}{2 t^{1/2}} \,\bigg|_{t_f}
= \frac{a(t_f)}{2t_f}. \label{a2der}
\end{align} 
By first expressing the Hubble parameter $H$ in terms of $N$, the number of inflationary $e$-folds
of expansion, and then equating derivatives, we find
\begin{align}
H &= \frac{N}{t_f - t_i} \label{Ndefn} \\
N &= \frac{t_f - t_i}{2t_f} = \frac{1}{2} - \frac{t_i}{2t_f}.
\end{align}
Continuity of the derivatives requires that $N \le \frac{1}{2}$, or else the time at the beginning of
inflation is less than zero. Although much research into inflation has produced a wide range of proposed
values for $N$, this result is particularly problematic. If taken literally, it would eliminate
inflation as a solution to the kinds of problems the theory was designed to solve.

\subsection{Quantifying the Discontinuity}
\label{Quantifying the Discontinuity}

Although early universe estimates are themselves quite problematic because of the uncertainty in the
values of basic parameters, using reasonable values can provide some insight into the mathematical
relationship between the scale factors in different periods of cosmological evolution. We can estimate a
value for $a(t_f)$ by taking advantage of the inverse relation between the scale factor and temperature,
in conjunction with estimates of temperature then and now, $T(t_f)$ and $T(t_0)$, respectively,
\begin{equation}
\label{atemp}
a(t_f) = \frac{T(t_0)}{T(t_f)} \approx
\frac{2.73 \, \text{K}}{1.16\times 10^{29} \, \text{K}}\approx 10^{-29},
\end{equation}
since by convention, $a(t_{0})=1$.
The temperature of the CMB today is $T(t_0) \approx 2.73$ K~\cite{fixsen2009}, and $T(t_f)$ corresponds to
the temperature equivalent to the value of $H$ for a universe that supports the Standard Model,
which is $H \approx 10^{16}$ GeV~\cite{weinberg2009}.

\begin{figure}[ht!]
    \centering
    \subfloat[\centering]{{\includegraphics[width=7cm]{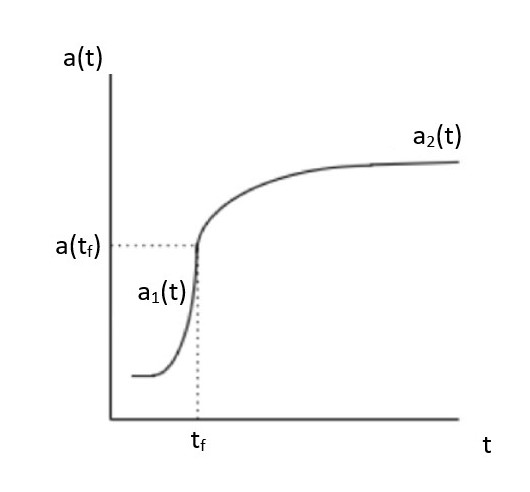} }}%
    \qquad
    \subfloat[\centering]{{\includegraphics[width=7cm]{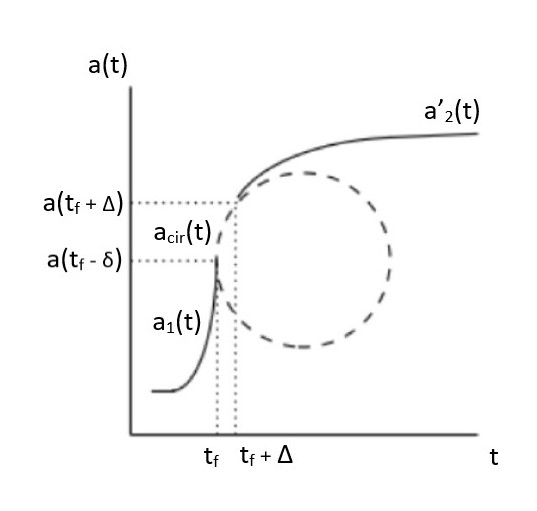} }}%
		\caption{(a) The graph depicts the discontinuity between inflationary and radiation era scale 
		factors at the end of inflation. (b) The circular arc defined by the transitional scale factor
		$a_{\text{cir}}(t)$ intersects tangentially with the inflationary scale factor $a_1(t)$ and the
		now-displaced radiation era scale factor noted with a prime, $a^\prime_2(t)$. The parameter
		$\Delta$ is the time period from the end of inflation until the time of continuity between
		$a_{\text{cir}}(t)$ and $a^{\prime}_2 (t)$. The $a$ functions are not to scale.}%
    \label{sidebyside}
\end{figure}

Next we compare slopes at the end of inflation. The inflationary slope is
\begin{equation} 
\dot a_1 (t_f) \approx 10^{-13} \, \text{GeV} \approx 10^{11} \, \text{s}^{-1}.
\end{equation} 
For the radiation-era derivative, after solving eq.~(\ref{Ndefn}) for $t_f$ and substituting it into
eq.~(\ref{a2der}), we have
\begin{equation}
\dot a_2 (t_f) = \frac{a(t_f)H}{2(N + Ht_i)} = \frac{\dot a_1 (t_f)}{2(N + Ht_i)}.
\end{equation} 
The estimate by Liddle and Lyth that inflation began at $t_i = 10^{-42 }$ s~\cite{LiddleLyth2006} leads to
$Ht_i \approx 0.02$. A reasonable assumption is that $N \approx 60$~\cite{linde2007,lyth2007,riotto2017},
which results in a measure of the discontinuity. The time derivative of the radiation-era scale factor is
approximately $\frac{1}{120}$ of the derivative of the inflationary scale factor. Graph~(a) of
figure~\ref{sidebyside} shows this change in the growth behavior qualitatively.
We also note that these values, $H = 10^{16}$ GeV and $N = 60$, yield an estimate for the
duration of inflation without the need to specify $t_i$ or $t_f$: 
\begin{equation}\label{dur}
\Delta t = \frac{N}{H} \approx 4 \times 10^{-39} \, \text{s}.
\end{equation}

\section{The Transition}
\label{The Transition}

What kind of transitional function could provide continuity between the two period scale factors? At the
end of inflation, the slope begins to decline. For simplicity, we require a steadily declining slope with
no regions in which the universe undergoes contraction. A properly chosen intermediate power law would
conform to our requirement, which we explore in section~\ref{Power Law Transitions}. Initially, as shown
in graph~(b) of
figure~\ref{sidebyside}, we shall use a circular arc $a_{\text{cir}}(t)$ to illustrate the geometry.
The arc lies tangent to $a_1(t)$ at the end of inflation and tangent to the now\nobreakdash-displaced
radiation-era scale factor noted with a prime, $a^{\prime}_2(t)$. The transitional duration $\Delta$
remains to be determined.

\begin{figure}[ht!]
\centering
\includegraphics[scale=0.40]{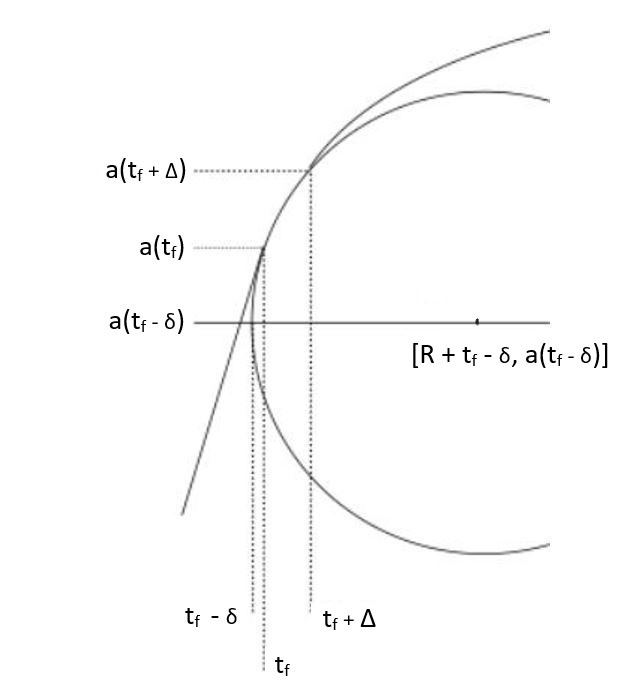}
\caption{Five unknown parameters characterize the two points of tangency of $a_{\text{cir}}(t)$
with the period scale factors (not to scale).}
\label{detailed_v1_w_margins_annot}
\end{figure}

In the more detailed view of figure~\ref{detailed_v1_w_margins_annot}, we see five unknown variables:
\begin{itemize}
\addtolength\itemsep{-2.9mm}
\item $R$ --- the radius of the circular arc
\item $\Delta$ --- the time between $t_f$ and the tangent point at which the circular arc $a_{\text{cir}}(t)$
meets the displaced radiation era $a^{\prime}_2(t)$
\item $a(t_f-\delta)$ --- the $a$-axis value at $t_f - \delta$, aligned with the center of the arc
\item $\delta$ --- the measure of the $t$-axis displacement corresponding to the difference between
$a(t_f)$ and  $a(t_f-\delta)$
\item $a (t_f + \Delta)$ --- the scale factor at $t_f + \Delta$, the $t$-axis point of
tangency for $a_{cir}(t)$ and $a^{\prime}_2(t)$.
\end{itemize}
The transitional function is just the equation of the circle
\begin{equation}
a_{\text{cir}}(t) = a(t_f-\delta) + \sqrt{R^2 - [t - (R + t_f -\delta)]^2}.
\end{equation}
If $a(t_f)$ and  $a(t_f-\delta)$ were to coincide, we would be introducing another discontinuity into the
model, the change from the inflationary slope to the infinite slope of the circular arc. With the
displacement $\delta$, the edge of the circular arc lies earlier on the timeline than $t_f$, and $\delta$
sets that duration. We note that $a(t_f-\delta)$ is therefore never the physical value of the scale factor
and so has no direct effect on the expansion of space.

We establish smoothness by equating the scale factors and their derivatives at the tangent points. Thus,
we have five unknown parameters in the four matching conditions. In
section~\ref{Continuity of the Equation of State}, we invoke a fifth equation to specify the model completely.

\section{Power Law Transitions}
\label{Power Law Transitions}

Imposing a circular arc is an unreasonably strict condition to use to define an interpolating function.
We shall continue now by exploring more general power-law solutions for $a(t)$ in the transition between
the period scale factors. A properly chosen power-law section would provide continuity but would also
potentially lengthen one or the other period, depending on its orientation. A power law with $n < 1$
has essentially no impact on the length of the radiation era. For a power law with $n > 1$, a very small
increase in the inflationary period could have a substantial impact on the scale factor, as discussed
further in section~\ref{Power Law with $n > 1$}.

\subsection{Interpolating Power Laws with $n < 1$}
\label{Power Law with $n < 1$}

The power law takes the form
\begin{equation}
a_p (t) = a_p(t_f - \delta) + D \, [t - (t_f - \delta)]^n.
\end{equation} 
We use the notation $a_p(t)$ for power laws with $n < 1$. The subscript $p$ denotes the representative
parabola for $n = \frac{1}{2}$, which opens to the right and has a horizontal axis. The unknown coefficient
$D$ is the analog of the unknown radius of the circular arc. Additional unknown parameters,
$a_p(t_f-\delta)$, $\delta$, $\Delta$, and $a(t_f + \Delta)$, correspond to the parameters displayed in
figure~\ref{detailed_v1_w_margins_annot} for the circular arc. We analyze the continuity of the forms
of the scale factor at the two points of tangency.

\subsubsection{Smoothness at $t_f$}
\label{Continuity at $t_f$}

For the first matching condition---continuity of $a(t)$---at $t_f$, $a_p (t) \big|_{t_f} = a(t_f)$ implies 
\begin{align}
D &= \frac{a(t_f) - a_p(t_f-\delta)}{\delta^n} \\
a_p (t) &= a_p(t_f-\delta) + \frac{a(t_f) - a_p(t_f-\delta)}{\delta^n} \, [t - (t_f - \delta)]^n.
\end{align}
The second matching condition at $t_f$ equates the time derivatives, generating an expression for the
$a$-axis vertex coordinate:
\begin{align}
\dot a_p (t) \big|_{t_f} &= \dot a_1(t) \big|_{t_f} = T(t_0) \\
\dot a_p(t)\big|_{t_f} &= n \bigg[ \frac{a(t_f) - a(t_f) - \delta}{\delta^n} \bigg] \delta^{n-1} \\
a_p(t_f-\delta) &= a(t_f) - \frac{T(t_0)\delta}{n}, \label{adisp}
\end{align}
where we have used the relations in eqs. (\ref{eq-a1der}) and (\ref{atemp}) to express these in terms of the
current temperature $T(t_{0})$, since $t=t_{0}$ also provides the calibration scale for $a$.
With the vertex coordinate from eq.~(\ref{adisp}), the scale factor is
\begin{equation}
\label{horpar}
a_p (t) = a(t_f) - \frac{T(t_0)\delta}{n} + \frac{T(t_0)}{n \delta^{n-1}} \, [t - (t_f - \delta)]^n.
\end{equation}

\subsubsection{Smoothness at $t_f + \Delta$}
\label{Continuity at tf plus Delta}

The interpolating transition we have imposed between the end of inflation and the beginning of the
radiation era shifts the eq.~(\ref{radera}) scale factor according to
\begin{equation}
a^{\prime}_2(t) = a(t_f + \Delta)\sqrt{\frac{t}{t_f + \Delta}},
\end{equation}
where we use the prime to distinguish this shifted expression. The vertex of the radiation-era $t^{1/2}$
scale factor remains at $(t=0,a=0)$. The third matching condition, in which the interpolating power-law
equals $a^{\prime}_2(t)$ at the point of tangency, yields the noninformative solution
\begin{equation}
a^{\prime}_2(t)\big|_{t_f + \Delta} = a_p(t_f + \Delta).
\end{equation}

However, the final smoothness condition equates the derivatives of the scale factors $a^{\prime}_2(t)$ and
$a_p (t)$ at $t_f + \Delta$, so we have
\begin{align}
\dot a_p (t) \big|_{t_f + \Delta} &= \dot a^{\prime}_2 (t) \big|_{t_f + \Delta} \\
\dot a^{\prime}_2 (t) \big|_{t_f + \Delta} &= \frac{a_p(t_f + \Delta)}{2(t_f + \Delta)} \\
\dot a_p (t) \big|_{t_f + \Delta} &= \frac{T(t_0)}{\delta^{n-1}} \,
[t - (t_f - \delta)]^{n-1} \bigg|_{t_f + \Delta} \\
&= \frac{T(t_0)}{\delta^{n-1}} \, (\Delta + \delta)^{n-1}. 
\end{align}
Thus we ultimately arrive at the condition,
\begin{align}
\frac{T(t_0)}{\delta^{n-1}} \, (\Delta + \delta)^{n-1} &= \frac{a_p(t_f + \Delta)}{2(t_f + \Delta)} \\
\frac{T(t_0)}{\delta^{n-1}} \, (\Delta + \delta)^{n-1} &= \frac{1}{2(t_f + \Delta)}
\bigg[ a(t_f) - \frac{T(t_0)\delta}{n} + \frac{T(t_0)}{n \delta^{n-1}} \, (\Delta + \delta)^n \bigg].
\end{align}
The formalism leaves us with the need to fix $\Delta$ to evaluate the model; $\Delta$ and $a (t_f + \Delta)$
are physical but as yet unknown parameters. The others are mathematical constructs with no direct
physical meanings. After substituting the early universe parameter values assumed in
section~\ref{Quantifying the Discontinuity}, we find, for example, the solution for $n=\frac{1}{2}$ at
$\Delta = 10^{-35}$ s 
\begin{equation}
\label{deltaDelta}
\delta \approx 2.65 \times 10^{-6} \Delta
\end{equation}
(evaluated using Maple).
Table~\ref{curve_table} lists additional values of $\delta$ for a sample set of transition durations
$\Delta$. The purpose of having three significant figures listed in the table is to illustrate the
relationship between the displacement and any changes to the transition scale.

We have accomplished the objective of parameterizing the transition from the inflationary scale
factor with a family of power-law scale factors that ensure sufficient smoothness to have continuity of $a$
and its first time derivative---although we have not imposed the condition of
continuity on any higher-order derivatives. The slope of the inflation-era scale factor grows at a rate on
the order of the Hubble parameter, and a requirement of continuity on the second derivative would
effectively extend inflation into the subsequent period, rather than marking the physical end of inflation
as the point at which the second derivative becomes negative.

\subsection{Power Laws with $n > 1$}
\label{Power Law with $n > 1$}

\begin{figure}[ht!]
\centering
\includegraphics[scale=0.60]{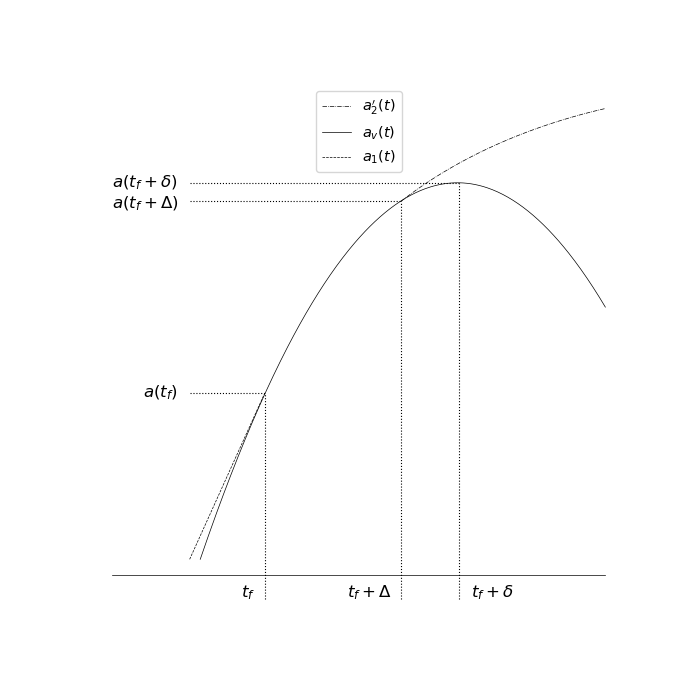}
\caption{The displacement $\delta$ necessarily places the vertex of the representative inverted parabola
with power law index $n = 2$ later than the end of inflation at $t_f$ and the tangency point at
$t_f + \Delta$ (not to scale). Increasing the duration over which the the interpolating scale factor
applies also shifts the vertex similarly.}
\label{avimage}
\end{figure}

To continue the study of alternative transitions, we now examine power laws with $n > 1$, containing
unknown parameters analogous to those analyzed in the previous section. The scale factor formula is
\begin{equation}
a_v (t) = a_v(t_f + \delta) + E \, [t - (t_f + \delta)]^n.
\end{equation}
The subscript $v$ denotes the representative inverted parabola for $n = 2$ with the vertical axis
parallel to the $a$-axis. The displacement of the vertex from $a(t_f)$ now places $\delta$ at a
time later than the tangent point at $t_f + \Delta$, as shown in figure~\ref{avimage}.

Repeating the analysis of the matching conditions at $t_f$ and $t_f + \Delta$ yields the scale factor  
\begin{equation}\label{verpar}
a_v (t) = a(t_f) + \frac{T(t_0)\delta}{n} + \frac{T(t_0)}{n(-\delta)^{n-1}} [t - (t_f + \delta)]^n
\end{equation}
and a fourth matching condition
\begin{equation}
\frac{T(t_0)}{(-\delta)^{n-1}} \, (\Delta - \delta)^{n-1} = \frac{1}{2(t_f + \Delta)}
\bigg[ a(t_f) + \frac{T(t_0)\delta}{n} + \frac{T(t_0)}{n (-\delta)^{n-1}} \, (\Delta - \delta)^n \bigg].
\end{equation}
As table \ref{curve_table} details for a sample set of transition durations, the
vertex displacement $\delta$ for a power
law with $n =2$ and $\Delta = 10^{-35}$ s is almost six orders of magnitude farther from the point of
tangency than that of the power law with $n=\frac{1}{2}$. The power law $a_p(t) \propto t^{1/2}$ can
establish continuity with the slope of the inflationary scale factor with such a minute displacement,
because the power law has an infinite slope at the vertex. However, the power law ${a_v(t) \propto t^2}$
has no such infinite slope, and the difficulty of establishing continuity with the large slope at the end
of inflation, $\sim 10^{11} \, \text{s}^{-1}$, manifests itself in the displacement being many orders
of magnitude greater than that of $a_p(t) \propto t^{1/2}$.

\begin{table}[H]\centering  
\begin{tabular}{l@{}ccccccc@{}}\toprule  
& Transition & \phantom{a} & $n$ & \phantom{a}& $\Delta \, \text{(s)}$ & \phantom{a}
& $\delta \, \text{(s)}$ \\ \midrule  
& $a_p(t)$ & \phantom{a} & $\frac{1}{2}$ & \phantom{a} & $10^{-35}$ & \phantom{a} & $2.65 \times 10^{-41}$ \\  
& \phantom{a} & \phantom{a} & \phantom{a} & \phantom{a} & $10^{-33}$ & \phantom{a} & $3.22 \times 10^{-41}$ \\ 
& \phantom{a} & \phantom{a} & \phantom{a} & \phantom{a} & $10^{-30}$ & \phantom{a} & $3.29 \times 10^{-41}$ \\ 
& \phantom{a} & \phantom{a} & \phantom{a} & \phantom{a} & $10^{-22}$ & \phantom{a} & $3.29 \times 10^{-41}$ \\ 
& $a_v(t)$ & \phantom{a} & $2$ & \phantom{a} & $10^{-35}$ & \phantom{a} & $1.50 \times 10^{-35}$ \\ 
& \phantom{a} & \phantom{a} & \phantom{a} & \phantom{a} & $10^{-22}$ & \phantom{a} & $1.50 \times 10^{-22}$ \\ 
\bottomrule  
\end{tabular}
\caption{The vertex displacements $\delta$ for power laws $a(t) \propto t^{1/2}$ and $t^2$ for a sample set of
transition durations $\Delta$ between the end of inflation and the beginning of the radiation era.
Increasing the duration $\Delta$ for the power law $n = \frac{1}{2}$ tends to set the displacement of the
vertex. However, because of the difficulty of establishing continuity with the inflationary slope at $t_f$,
the displacement $\delta$ for the $n = 2$ interpolator is many orders of magnitude greater. With its vertex
located later on the timeline than $t_f$, the table shows that increasing the transition has the effect
of shifting the vertex of $a_v(t)$ farther away from the end of inflation.}
\label{curve_table}  
\end{table} 

\begin{figure}[ht!]
\centering
\includegraphics[scale=0.90]{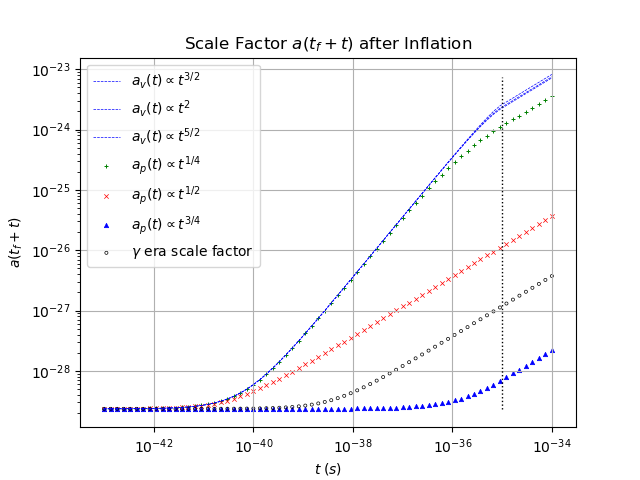}
\caption{The three $a_v(t)$ scale factors with power laws $t^{3/2}$, $t^2$, and $t^{5/2}$
essentially overlay each other, and the $a_p(t)$ scale factor proportional to $t^{1/4}$ approaches
those of the power laws with $n > 1$. The graph also shows the unsmoothed scale factor defined by
eq.~(\ref{radera}), which exceeds that of $a_p(t)$ at $t^{3/4}$. The $t$-axis timeline begins at
$t = 10^{-43}$ s after inflation terminates, while the vertical dotted line at $t = 10^{-35}$ s marks the
nominal start of the radiation era.}
\label{scale factors -35}
\end{figure}

Figure~\ref{scale factors -35} displays the transitional scale factors of eqs. (\ref{horpar}) and
(\ref{verpar}), rescaled by a translation of the $t$-axis $t \rightarrow t^\prime = t_f + t$.
The timeline starts at the arbitrarily small initial value $t = 10^{-43}$ s. The graphs represent power laws
for $n < 1$ and $n > 1$ with a representative sample of powers. 
However, the graphs of the scale factors themselves offer somewhat limited insight into the evaluation
of the quality of the interpolating functions. For that, we now look instead to graphs of the Hubble
parameter, 
\begin{equation}
H(t_f + t) = \frac{\dot a(t_f + t)}{a(t_f + t)}.
\end{equation}
In the transitional Hubble parameters below, $H_{\text{in}}$ is the constant inflationary Hubble parameter,
taken to be $10^{16}$ GeV. For the two classes of power laws, we find
\begin{align}
H_p(t_f + t) &= \text{ \large $\frac{\left(1 + \frac{t}{\delta} \right)^{n-1}}
{\frac{1}{H_{\text{in}}} - \frac{\delta}{n} + \frac{\delta}{n}\left( 1 + \frac{t}{\delta} \right)^n}$ }
\label{Hp} \\
H_v(t_f + t) &=  \text{ \large $\frac{\left(1 - \frac{t}{\delta} \right)^{n-1}}
{\frac{1}{H_{\text{in}}} + \frac{\delta}{n} - \frac{\delta}{n}\left( 1 - \frac{t}{\delta} \right)^n}$ }.
\label{Hv}
\end{align}

\begin{figure}[ht!]
\centering
\includegraphics[scale=0.90]{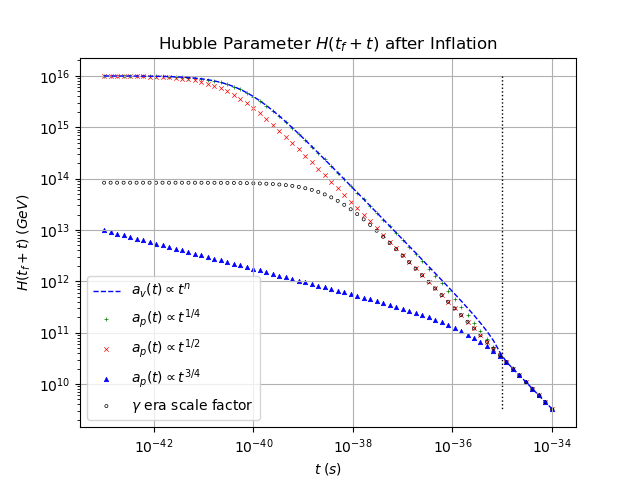}
\caption{The Hubble parameters corresponding to the scale factors shown in figure~\ref{scale factors -35}.
Once again, the graphs based on the scale factors $a_v(t)$ for $t^{3/2}$, $ t^2$, and $ t^{5/2}$
essentially overlay each other. The discontinuity of the radiation-era Hubble parameter with
$H_{\text{in}}$ exceeds two orders of magnitude. The graph based on scale factor $a_p(t)$ for $t^{3/4}$
fails to display asymptotic behavior with $H_{\text{in}}$ at small times,
because the timeline has the same $t$-axis translation and does not start at $t_f$.
Again the vertical dotted line marks the start of the radiation era at $t = 10^{-35}$ s.}
\label{combined H -35}
\end{figure}

In figure~\ref{combined H -35}, showing graphs of eqs.~(\ref{Hp}) and (\ref{Hv}), we note the requirement
of smoothness at the beginning of the radiation era causes abrupt shifts downward and upward as
$t \rightarrow \Delta$ for interpolating scale factors not proportional to $t^{1/2}$. We note that the
power laws with $n > 1$ that overlay each other in both figures~\ref{scale factors -35} and
\ref{combined H -35} must exhibit the shift downward to establish continuity with the radiation era Hubble
parameter. Unable to justify a physical basis for this behavior, we shall move forward in our analysis by
eliminating these power laws as valid interpolating functions and focus on more specifically determining
workable interpolating functions with $n < 1$. We also note that as $n \rightarrow \frac{1}{2}$ from above
or below, the scale factor $a_p(t)$ transforms more seamlessly into the radiation era. So at this stage,
we expect that the most suitable interpolating functions will correspond to the index value $n=\frac{1}{2}$,
or something close to that. The power laws for the interpolating region and the subsequent radiation-dominated
era are both horizontally-opening parabolas (or nearly so), which differ principally in their vertex placements
and radii of curvature.

\section{Additional Constraints}
\label{Additional Constraints}

\subsection{The Equation of State}
\label{The Equation of State}

\begin{figure}[ht!]
\centering
\includegraphics[scale=0.80]{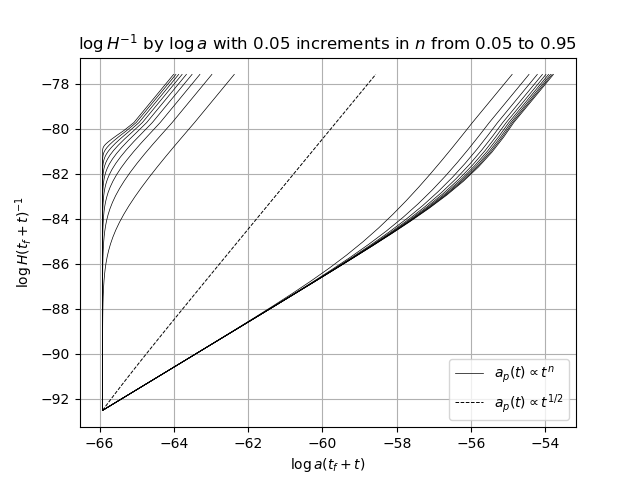}
\caption{The slopes of the graphs equal the parameter $\epsilon_H = \frac{d \, \log (H^{-1})}{d \log \, a}.$}%
\label{loganov increments pt05}
\end{figure}

We continue with the evaluation of the usefulness of the possible interpolations by considering a
parameter $\epsilon_H$, which is an alternative to the equation-of-state parameter $\omega$ that satisfies
$p=\omega\rho$, according to
\begin{equation}
\epsilon_H = \frac{d \log (H^{-1})}{d \log a} = \frac{3}{2}(1 + \omega).
\end{equation}
Following a graphical technique described by Kaloian Lozanov~\cite{lozanov2019}, we shall interpret the
formula as the slope in a plot of the evolution of the scale factor from inflation through reheating, 
matter domination, and finally the dark-energy-dominated era. Appendix~\ref{eHworking} provides more
information about this expression. Figure~\ref{loganov increments pt05} shows our version of the
Lozanov graphical approach for the power law $a_p(t) \propto t^n$, with $n$ ranging from $0.05$ to
$0.95$. Table~\ref{epH} lists statistics for some of the graphed power laws, as well as smaller and larger 
values of $n$.

Aside from the footnoted observations in table~\ref{epH}, we note a further
curious feature of figure~\ref{loganov increments pt05}. The graphs at the upper and lower extremes
of $n$ display almost cusp-like changes of slope at the tangent point between the transition and the
radiation-dominated era at $t = t_f + \Delta$. Between inflation and the start of the radiation-dominated era,
$\epsilon_H$ changes from a value much greater than 2 to less than 2 for $n > \frac{1}{2}$. Conversely, for
$n < \frac{1}{2}$, the parameter starts less than 2 and then becomes greater than 2. With our expectation
that $\epsilon_H = 2$, only the power law $n = \frac{1}{2}$ (the same power law index as in the radiation
era itself) appears able to transition seamlessly to the
radiation era, which suggests that all powers except $n = \frac{1}{2}$ result in a cusp in the
evolution of $\epsilon_H$. This motivates a closer inspection.

\begin{table}[H]\centering  
\begin{tabular}{l@{}lllllll@{}}\toprule  
& $n$ & \phantom{a} & $\delta \, \text{(s)}$ & \phantom{a}& $\epsilon_H$ & \phantom{a} & $\omega$ \\ \midrule  
& 0.002 & \phantom{a} & $2.53 \times 10^{-36}$ & \phantom{a} & $2.59$ & \phantom{a} & $0.73^{\, \text{(b)}}$ \\  
& 0.25 & \phantom{a} & $9.55 \times 10^{-37}$ & \phantom{a} & $2.37$ & \phantom{a} & $0.58$ \\  
& 0.50 & \phantom{a} & $2.65 \times 10^{-41}$ & \phantom{a} & $2.00^{\, \text{(a)}}$ & \phantom{a} & $0.33$ \\  
& 0.75 & \phantom{a} & $9.46 \times 10^{-56}$ & \phantom{a} & $1.50$ & \phantom{a} & $0.00^{\, \text{(c)}}$ \\ 
& 0.98 & \phantom{a} & $2.31 \times 10^{-294}$ & \phantom{a} & $1.04$ & \phantom{a} & $-0.31^{\, \text{(d)}}$ \\  
\bottomrule  
\end{tabular}
\caption{The displacement $\delta$, parameter $\epsilon_H$, and equation-of-state parameter
$\omega = \frac{2}{3} \epsilon_H - 1$ for power laws $a_p(t) \propto t^n$,
with values of $n$ ranging between 0 and 1.
\newline
(a) Computation sets this value more precisely at $\approx 2.00039$. After interpolation using the linear
relation associated with eq.~(\ref{eHeq}), the expected $\epsilon_H = 2$ for a radiation-dominated
scale factor occurs at $n \approx 0.5002$. 
\newline
(b) Values of $\epsilon_H \rightarrow 2.6^{-}$ and $\omega \approx 0.73$ signify unphysical, exotic
tachyon-like particles with velocities greater than the speed of light, which
section~\ref{Speed of Sound Constraints} discusses in detail.
\newline
(c) For $\epsilon_H \approx 1.50$ and $\omega \approx 0.00$, we have a transition from inflation to an
equation of state that would be consistent with a matter-dominated universe. We take up consideration of
the single-component matter-dominated universe in section \ref{Continuity of the Equation of State}.
\newline
(d) As $\epsilon_H \rightarrow 1.0^{+}$ and $\omega \rightarrow -\frac{1}{3}$, the scale factor remains
inflationary, effectively eliminating the transition.}
\label{epH}    
\end{table}

\begin{figure}[ht!]
\centering
\includegraphics[scale=0.80]{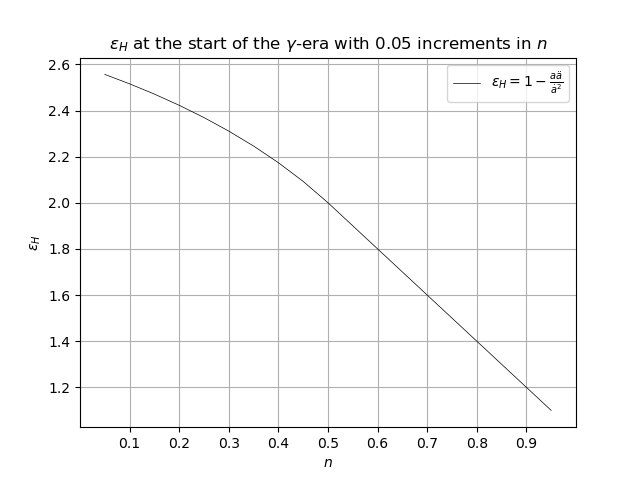}
\caption{The parameter $\epsilon_H = -\frac{\dot{H}}{H^2}$ with power law $a_p(t) \propto t^n$ for
$n <1$ at the start of the radiation era, with the transition period $\Delta = 10^{-35}$ s.}%
\label{cusp graph}
\end{figure}

Figure~\ref{cusp graph} plots $\epsilon_{H}$ versus the power law index $n$ at time $(t_f + \Delta)$:
\begin{equation}\label{eHeq}
\epsilon_H = 1 + (1 - n) \frac{a(t_f) - \frac{T(t_0)\delta}{n} +
\frac{T(t_0)\delta}{n}\left(1 + \frac{\Delta}{\delta} \right)^{n}}{T(t_0)\delta
\left(1 + \frac{\Delta}{\delta} \right)^n}.
\end{equation}
For $n > \frac{1}{2}$, the negative term in the numerator of eq.~(\ref{eHeq}) is small relative to the other
terms appearing in the fraction, which are on the order of $a(t_f)$ and essentially cancel with the factor of
the same characteristic size in the denominator. The linear relation $\epsilon_H-1 \propto (1-n)$ remains,
as the graph and table~\ref{epH} show. In contrast, for $n < \frac{1}{2}$,
the scale factor $a(t_f)$ is small relative to the other terms in the fraction. As $n \rightarrow 0$,
the fraction and $(1 - n)$ are both increasing, and $\epsilon_H$ increases to approximately $2.6$.

Figure~\ref{cusp graph} aligns with the possibility raised by figure~\ref{scale factors -35}
that only $n = \frac{1}{2}$ results in a transition to the radiation era without an awkward,
cusp-like feature---whose very presence would seem to be contrary to the dictum we have adopted of modeling
the transitions in a smooth fashion. However, precise calculations consistently indicate
that a slightly different $n \approx 0.5002$ actually produces the seamless transition. In the same way
as the abrupt shift that we cannot explain in the graph of the Hubble parameters tends to disqualify all
power laws except $n=\frac{1}{2}$, the cusp-like features again appear likely to signify unphysical,
unexplained behavior. However, before we attempt to resolve these conflicts, we shall review additional
constraints on the equation of state, starting with constraints related to the speed of sound.

\subsection{Speed of Sound Constraints}
\label{Speed of Sound Constraints}

Another tool for evaluating the interpolators is the application of constraints on the speed of sound
to the equation of state. A speed of sound less than zero or greater than the speed of light would
violate stability or causality, respectively~\cite{lindblom2018, ellis2007}. Stability requires that the
speed must be real; imaginary phase speeds would correspond to imaginary frequencies, or modes that grow
exponentially with time. At the other end, special relativity imposes the standard limitation that information
carried by arbitrary quanta cannot propagate faster than the speed of light $c=1$ in vacuum. Thus, we expect
the sound speed of the transition waves to obey inequalities
\begin{equation}\label{sslim}
0 \le v_s^2 \le 1.
\end{equation}

We assume, as is standard, that the inflaton wave oscillations are fast and thus
adiabatic, so that a
passing wave brings about temperature changes without conductive heat transfer. The thermodynamic
behavior is reversible, and so the entropy per unit mass is constant~\cite{thorne2017} as an inflaton wave
passes through. Pressure ${p = p(s,\rho)}$ becomes a function of the density $\rho$ only, ${p = p(\rho)}$.
We also make the assumption that the inflaton condensate at the end of inflation is a perfect fluid,
allowing us to apply a linear, single-component equation of state, $p = \omega\rho$, expressing the
dependence of pressure on density in terms of a $\rho$-independent equation-of-state parameter $\omega$.
This environment yields a sound speed
\begin{equation}
v_{s}^{2}=\frac{1}{d\rho/dp}=\frac{dp}{d\rho}=\omega,
\end{equation}
and using $\omega = \frac{2}{3}\epsilon_H - 1$, the condition $0 \le \omega \le 1$ implies
$\frac{3}{2} \le \epsilon_H \le 3$. At the precise end of inflation, $\Delta = 0$ in eq.~(\ref{eHeq}),
leaving
\begin{equation}
\epsilon_H = 1 + (1 - n) \frac{a(t_f)}{T(t_0) \delta} = 1 + \frac{(1 - n)}{H \delta}.
\end{equation}

\begin{figure}[ht!]
\centering
\includegraphics[scale=0.6750]{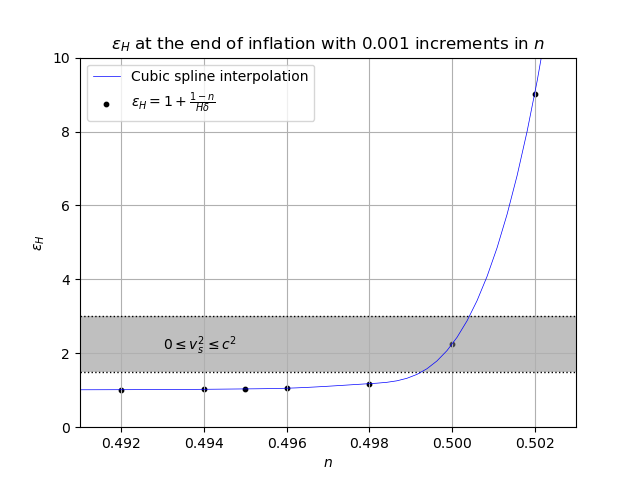}
\caption{Values of the parameter $\epsilon_H = 1 + \frac{(1 - n)}{H \delta}$ at the end of inflation,
$t_f \approx 4 \times 10^{-39}$ s, for $a_p(t) \propto t^n$ with $n \approx \frac{1}{2}$. The gray band
depicts the permissible values of $\epsilon_H$, $\frac{3}{2} \le \epsilon_{H} \le 3$, subject to the assumption
that the inflaton condensate at the end of inflation is a single-component perfect fluid with equation of
state $p = \omega\rho$.}%
\label{TD constraints drill down}
\end{figure}

Figure~\ref{TD constraints drill down} displays the effect of enforcing the sound speed restrictions from
eq.~(\ref{sslim}) on $\epsilon_H$; these conditions severely restrict the permissible range of power laws
indices. The line plot is a Python cubic spline interpolation of Maple-generated solutions for the parameter
$\epsilon_{H}$ from eq.~(\ref{eHeq}) in 0.001 increments around $n = \frac{1}{2}$. The section of the spline
interpolation within the gray horizontal band contains valid values of $\epsilon_{H}$, corresponding to
interpolating theories with stable, causal sound speeds. Reading off the graph, we see the permissible
physical power law band for the continuous function transitioning from the end of inflation to the radiation
era lies approximately between $0.4990 < n < 0.5005$. This narrow band is consistent with the
large separation in figure~\ref{loganov increments pt05} between the power law $n = \frac{1}{2}$ and the
closely adjacent powers, $n = 0.45$ and $n = 0.55$.

\subsection{Continuity of the Equation of State}
\label{Continuity of the Equation of State}

Although we have found restrictions on the allowed range of power law values,
a tension between the expected and derived equations of state for $n = \frac{1}{2}$ actually remains.
We do not find that $\epsilon_{H}=2$ corresponds exactly to $n=\frac{1}{2}$.
We instead recall the expected $\epsilon_H = 2$ was associated with $n \approx 0.5002$ reported
in table~\ref{epH}, and we now seek to understand the reason for the slight discrepancies between these
values and the derived $\epsilon_H \approx 2.00039$ that corresponds to the exact $n = \frac{1}{2}$.

Comparison of the scale factor in eqs.~(\ref{adisp}) and (\ref{horpar}) of
section~\ref{Power Law with $n < 1$} with the first two terms in the numerator of the fraction in
the expression below,
\begin{equation}
\label{eHderived}
\epsilon_{H\,\text{derived}} = 1 + (1 - n) \frac{a(t_f) - \frac{T(t_0)\delta}{n}
+\frac{T(t_0)\delta}{n}\left(1 + \frac{\Delta}{\delta} \right)^{n}}{T(t_0)\delta
\left(1 + \frac{\Delta}{\delta} \right)^n},
\end{equation}
indicates that those two terms arise from the $a$-axis scale factor displacement
$a_p(t_f - \delta)$ of the interpolating function's vertex.  The third term in the numerator
represents the functional dependence of the scale factor on time. 

Table~\ref{nera} separates the difference between the expected and derived $\epsilon_H$ into the
relative contributions from the components of the numerator, $\epsilon_{H\,\text{displacement}}$
and $\epsilon_{H\,\text{time}}$. For comparison, we repeat the analysis for a single-component,
matter-dominated transition function establishing continuity between the end of inflation and a
matter-dominated era (that is, with $n=\frac{2}{3}$ power laws). We see the results are qualitatively the same. The displacement of the vertex of the scale factor along the $a$-axis is
responsible for the discrepancies.

\begin{table}[H]\centering  
\begin{tabular}{l@{}ccccccccccc@{}}\toprule  
& $n$ & \phantom{a} & $\delta$ (s) & \phantom{a}& $\epsilon_{H\,\text{expected}}$ & \phantom{a} 
& $\epsilon_{H\,\text{derived}}$ & \phantom{a} & $\epsilon_{H\,\text{displacement}}$ & \phantom{a}
& $\epsilon_{H\,\text{time}}$\\
\midrule  
& $\frac{1}{2}$ & \phantom{a} & $2.65 \times 10^{-41}$ & \phantom{a} & $2$ & \phantom{a} & $2.00039$
& \phantom{a} & $0.00039$ & \phantom{a} & $1.00000$ \\ [.15cm] 
& $\frac{2}{3}$ & \phantom{a} & $7.71 \times 10^{-42}$ & \phantom{a} & $1.5$ & \phantom{a} & $1.50020$
& \phantom{a} & $0.00020$& \phantom{a} & $0.50000$ \\ 
\bottomrule  
\end{tabular}
\caption{These parameters correspond to power laws $a_p(t)$ with indices $n=\frac{1}{2}$ and
$n=\frac{2}{3}$ transitioning to radiation-dominated and matter-dominated eras with scale factors
likewise proportional to $t^{1/2}$ and $t^{2/3}$.
The column $\epsilon_{H\,\text{derived}}$ reconstructs the parameter as the sum of 1 and the contribution
from the displacement and the time components. We conclude that the displacement causes the difference from
$\epsilon_{H\,\text{expected}}$.}
\label{nera}    
\end{table}

If not for the contribution of the vertex displacement, we would have seamless transitions of the
equation of state between the interpolating power laws and the radiation or matter eras. A first-order
phase transition at $(t_f + \Delta)$ might be responsible for the cusp, but we reason against that
possibility. The dynamics of the expansion of space at the tangent point undergoes no change. Prior to and
after $(t_f + \Delta)$, the power law index $n$ governing expansion remains approximately the same for
each single-component era. Also, the transition precedes the period of preheating described in
section~\ref{Effect of the Continuous Scale Factor} and subsequent thermalization, so that we expect
temperature to evolve smoothly at $t_f + \Delta$.

Instead, invoking continuity of the equation of state at $(t_f + \Delta)$ and noting
$\epsilon_{H\,\text{expected}} = \frac{1}{n}$ for a single-component universe, as appendix~\ref{singcomp}
shows, we solve eq.~(\ref{eHderived}) for the displacement $\delta$ and find
\begin{align}
\epsilon_{H\textit{expected}} = \frac{1}{n} &= 1 + (1 - n) \frac{a(t_f) - \frac{T(t_0)\delta}{n}
+ \frac{T(t_0)\delta}{n}\left(1 + \frac{\Delta}{\delta} \right)^{n}}{T(t_0)\delta
\left(1 + \frac{\Delta}{\delta} \right)^n} \\
 \frac{1}{n} &= \frac{a(t_f) - \frac{T(t_0)\delta}{n} + \frac{T(t_0)\delta}{n}
\left(1 + \frac{\Delta}{\delta} \right)^{n}}{T(t_0)\delta\left(1 + \frac{\Delta}{\delta} \right)^n}.
\end{align}
Substituting $a(t_f) = T(t_0)/H$ from eq.~(\ref{atemp}) with $H = H_{\text{in}}$ yields
\begin{equation}
\label{delta}
\delta = \frac{n}{H}.
\end{equation}
Analysis of eq.~(\ref{eHderived}) demonstrated that the displacement 
\begin{equation}
\epsilon_{H\,\text{displacement}} = a(t_f) - \frac{T(t_0)\delta}{n}
\end{equation}
caused the cusp-like feature. With the eq.~(\ref{delta}) result, and
recalling $a(t_f) = \frac{T(t_0)}{H}$, we instead have $\epsilon_{H\,\text{displacement}} = 0$, and the
bump on the curve is gone.

However, making the assumption of continuity of the equation of state at $t_f + \Delta$ destroys the
smoothness of the scale factor that we imposed at both $t_f + \Delta$ and $t_f$. So we must reexamine
the matching condition at $t_f + \Delta$; returning to the fourth matching condition and trying to solve
for $\Delta$, we see that
\begin{align}
\frac{T(t_0)}{\delta^{n-1}} \, (\Delta + \delta)^{n-1} &= \frac{1}{2(t_f + \Delta)}
\left[ a(t_f) - \frac{T(t_0)\delta}{n} + \frac{T(t_0)}{n \delta^{n-1}}\left(\Delta + \delta\right)^{n}
\right] \\
T(t_0)\left( 1 + \frac{\Delta}{\delta} \right)^{\!n-1} &= \frac{1}{2(t_f + \Delta)}
\left[ \frac{T(t_0)}{H} - \frac{T(t_0)\delta}{n} + \frac{T(t_0)\delta}{n}
\left( 1 + \frac{\Delta}{\delta} \right)^{\!n} \right] \\
0 &= \frac{1}{H} - \frac{\delta}{n} + \frac{\delta}{n}\left( 1 + \frac{\Delta}{\delta} \right)^{\!n}
- 2(t_f + \Delta) \left( 1 + \frac{\Delta}{\delta} \right)^{\!n-1}.
\end{align}
With $\delta = \frac{n}{H}$, this simplifies to
\begin{align}
2(t_f + \Delta) &= \frac{1}{H} \Big( 1 + \frac{H\Delta}{n} \Big) \\
\Delta &= \frac{\frac{1}{H} - 2t_f}{2 - \frac{1}{n}}.
\end{align}
So the transition period $\Delta$ is undefined for $n = \frac{1}{2}$, which invalidates the claim of
first-derivative smoothness imposed by the eq.~(\ref{deltaDelta}) parameters,
$\delta \approx 2.65 \times 10^{-6} \Delta$. For $n \approx 0.5002$, associated with $\epsilon_H = 2$,
the new formula's value of $\Delta$ is in fact less than zero. Since $\Delta$ is supposed to represent
the length of time over which the interpolating function applies, this value is manifestly unphysical.

Furthermore, substituting $\delta = n/H$ in the interpolating scale factor,
\begin{align}
a_p (t) &= a(t_f) - \frac{T(t_0)\delta}{n} + \frac{T(t_0)\delta}{n}
\left(1+ \frac{t - t_f}{\delta}\right)^n \\
&= \frac{T(t_0)\delta}{n} \, \left(1+ \frac{t - t_f}{\delta}\right)^n,
\end{align}
eliminates the $a$-axis displacement of $a_p(t)$.
We introduced the displacement of the power-law vertex in section~\ref{Continuity at $t_f$}
in order to enforce the smoothness condition at $t_f$, but this would be undone by the assumption of
exact continuity of the equation of state.

The Lozanov graphical approach to analyzing the equation of state suggests a range of power law indices,
$0.5000 \lesssim n \lesssim 0.5002$, are reasonable, and this is supported by the values the are permitted by
the speed-of-sound constraints, $0.4990 \lesssim n \lesssim 0.5005$. Having tried unsuccessfully to
establish continuity with the equation of state at $t_f + \Delta$, we now seek an explanation of the
discontinuity. If the transition represents a continuation and ultimately a termination of inflation,
a local discontinuity might result from a weak phase transition of unknown character. A second explanation
may be that a power law index not equal to 0.5 in the transition signals that the composition of the
universe is not strictly radiation-dominated as the transition ends, and so a single-component model is
not sufficient to describe the dynamics.

\section{Summary of Numerical Results}
\label{Results of the Numerical Analysis}

\begin{figure}
    \centering
    \subfloat[\centering]{{\includegraphics[scale=0.6750]{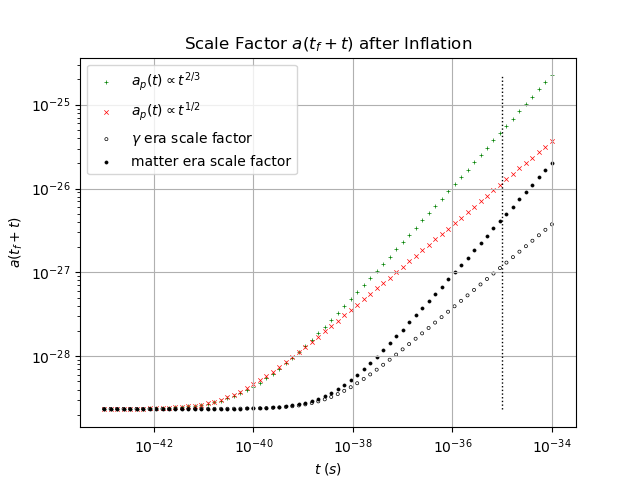} }}%
    \qquad
    \subfloat[\centering]{{\includegraphics[scale=0.6750]{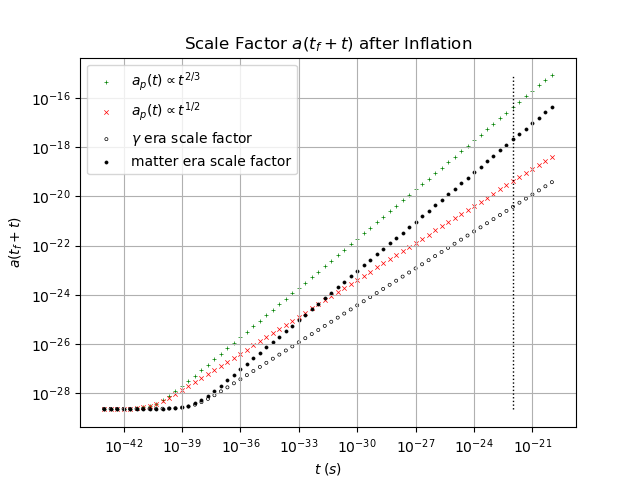} }}%
		\caption{The growth in scale factors for  single-component universes with smoothness enforced at
		$t_f + \Delta$ with $\Delta = 10^{-35}$ s and $10^{-22}$ s in (a) and (b), respectively.
		We note that the approximate order-of-magnitude increases in the power law scale factors
		occur at around $10^{-37}$ s in all cases.}%
		\label{sidebyside2}
\end{figure}

We have found that a transition function with power-law index $n$ can provide seamless
first-derivative smoothness over the period between the end of inflation and the development of a
single-component $n$-era universe, while obeying fundamental stability and causality constraints.
During the transition, the scale factor increases by approximately an order of magnitude more, compared with
what it would have been in a
model with a sharp cusp dividing the inflationary from post-inflationary functional forms; and that
additional accumulated expansion factor remains as time progresses. Figure~\ref{sidebyside2} shows the key
comparisons. 

\begin{figure}[ht!]
\centering
\includegraphics[scale=0.6750]{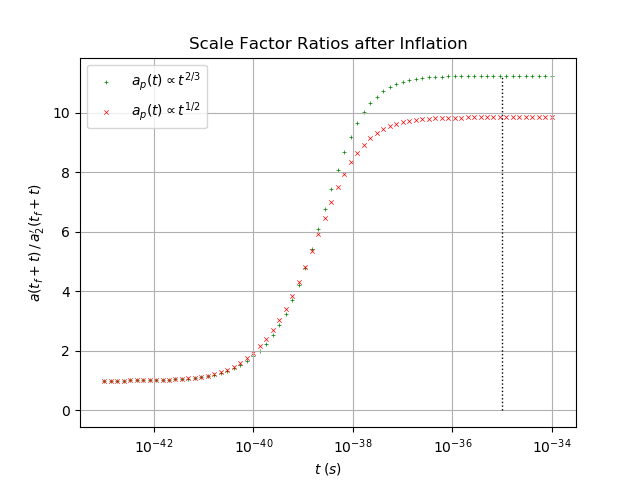}
\caption{The scale factor ratios $\frac{a_p(t)}{a^\prime_2(t)}$. At approximately $t = 10^{-37}$ s,
the ratios reach greater than $98\%$ of the asymptotic values of $9.8$ and $11.2$ for
$a_p(t) \propto t^{1/2}$ and $a_p(t) \propto t^{2/3}$, respectively.}
\label{scale factor ratios -35}
\end{figure}

Figure~\ref{scale factor ratios -35} depicts the increases toward asymptotic limits more clearly.
Both figures~\ref{sidebyside2} and~\ref{scale factor ratios -35} also reveal that these increases
occur primarily in the vicinity of $t=10^{-37} \, \text{s}$ and do not particularly depend
on the duration of the transition $\Delta$. Table~\ref{tableratios} contains further data, including
how much larger, relatively speaking, the universes with the smooth interpolations are than the models
without smoothing. The underlying numbers show that after
$10^{-34}$~s, the asymptotic values $9.8$ and $11.2$ have completely stabilized (to over
$12$-decimal-place precision). Even by $10^{-37}$~s, the increased expansion factors have already
grown to be within 2\% of their asymptotic values.

\begin{table}[H]
{\tabcolsep=0pt\def\arraystretch{1.3}}
\begin{center}
\begin{tabularx}{280pt}{c *8{>{\Centering}X}}\toprule
& \multicolumn{1}{l}\phantom{ }  & \multicolumn{3}{c}{$\;\;\;a_p(t) \propto t^{1/2}\;\;\;$} &
\multicolumn{1}{l}\phantom{ }   & \multicolumn{3}{c}{$a_p(t) \propto t^{2/3}$}
\tabularnewline \cmidrule(lr){3-5}\cmidrule(l){7-9}
  Time (s) & \phantom{abc} & Ratio & \phantom{abc} & {\%} & \phantom{abc} & Ratio & \phantom{abc} 
  & \%
\tabularnewline \midrule
  $10^{-34}$  & \phantom{abc} &  9.837 &  \phantom{abc} & 100 & \phantom{abc} & 11.227 & \phantom{abc}
  & 100 \tabularnewline \midrule\midrule
  $10^{-36}$  &  \phantom{abc} & 9.828 &  \phantom{abc} & 99.9 & \phantom{abc} & 11.217 & \phantom{abc}
  & 99.9 \tabularnewline
  $10^{-37}$  &  \phantom{abc} & 9.69 & \phantom{abc} & 98.5 & \phantom{abc} & 11.03 & \phantom{abc}
  & 98.3 \tabularnewline
  $10^{-38}$  &  \phantom{abc} & 8.4 & \phantom{abc} & 85.8 & \phantom{abc} & $\;\,9.4$ & \phantom{abc}
  & 83.3 \tabularnewline
  \bottomrule
\end{tabularx}
\captionof{table}{The ratios of $a_p(t)$ to $a^{\prime}_2(t)$ at different times during the
interpolation period. The percentages represent the degree to which
the ratios have approached the asymptotic values reached at $10^{-34}$ s.}
\label{tableratios}
\end{center}
\end{table}

We are left with the interesting result about what happens when we insert an interpolating
function after the end of inflation to smooth out the dynamics. Compared with the models with
discontinuous derivatives---signifying abrupt transitions between the inflationary period and a period
with a different equation of state---the total expansion of the scale factor is greater 
by about an order of magnitude (or between 2 and 3 $e$-folds). In a way, this is unsurprising, since
the interpolating function allows the inflationary expansion
to tail off a bit more gradually, and so the net result is always a larger universe at later times.
This kind of increase in the scale factor will form the basis for
our analysis of the effect of continuity in the numerical analysis going forward.

\section{The Smooth Scale Factor in the Preheating Model}
\label{Effect of the Continuous Scale Factor}

Having concluded that enforcing a smooth transition results in an order-of-magnitude increase in the
ultimate scale factor of the subsequent single-component universe, we shall now examine the effects of
this change on reheating, based on the preheating model of KLS~\cite{kls1997},
in which the inflaton couples to a second scalar field $\chi$ in the era following
after inflation, which is taken to be a matter-dominated universe.
We shall evaluate preheating effects using a smooth interpolating power law with
$n = \frac{2}{3}$, as described previously in section~\ref{Power Law with $n < 1$} as an example
of a power law with
$n < 1$. We compare our results to those of the KLS model, which employs the scale factor
$a(t) \approx a_{f}(t/t_{f})^{2/3}$ with a discontinuous slope. Our numerical analysis shows that the larger
scale factor in the smooth model decreases the $\chi$ occupation numbers $n_k$ and dilutes the total
number density $n_\chi$. The dilution arises
naturally out of the volume increase due to the greater expansion of space---although
the broad parametric resonance during preheating partially offsets the effect. Broad parametric
resonance involves all modes of the scalar field $\chi$ less than a specific maximum being involved in
quasi-resonant interactions with the inflaton, and it
causes an exponential increase in the number of $\chi$ particles created.

\subsection{Occupation Numbers}
\label{Occupation Numbers}

In this section and section~\ref{Number Density}, we briefly summarize the foundations of the detailed,
extensive case that KLS present in support of their theory. The Lagrange density for the scalar field $\chi$
coupled to the inflaton,
\begin{equation}
\mathcal{L}=-\frac{1}{2}\chi_{,\mu}\chi^{,\mu}-\frac{1}{2}m_\chi^2\chi^2 - \frac{1}{2}g^2\phi^2\chi^2,
\end{equation} 
in expanding space with vanishing mass parameter $m_\chi = 0$, generates the equation of motion
\begin{equation}
\ddot \chi_k + \frac{3 \dot a}{a}\dot\chi_k + \bigg( \frac{k^2}{a^2} +g^2 \phi^2 \bigg) \chi_k=0,
\end{equation}
where $k = \sqrt{\mathbf{k}^2}$, for the Fourier mode $\chi_k$ in momentum space. The inflaton at the end
of inflation is a coherently oscillating field of form $\phi(t) =\Phi(t)\sin(mt)$,
with amplitude envelope $\Phi(t) = \frac{M_{P}}{\sqrt{3\pi}mt}$~\cite{allahverdi2010}, so that
\begin{equation}\label{eompre}
\ddot \chi_k + \frac{3 \dot a}{a}\dot\chi_k +\left[\frac{k^2}{a^2} +g^2 \Phi^2(t)\sin^2(mt)\right]\chi_k=0.
\end{equation} 
In slow roll inflation, chaotic inflation, and other inflationary models in which the friction term
$3H\dot\phi$ in the equation of motion (\ref{eomft}) becomes negligible, the inflaton exhibits
sinusoidal oscillating behavior around $\phi = 0$. (Here the argument of the sine function has time $t$ in
units of $\text{m}^{-1}$, which the KLS model uses throughout.) The
appearance of the Planck mass $M_P$ in $\Phi(t)$ derives
from the Hubble parameter expressed in terms of the gravitational constant. The units of $k$ are m, and
the scale factor, normalized in the Robertson–Walker metric with $a(t_{0}) = 1$ today, remains dimensionless.

Broad parametric resonance consists of non-adiabatic oscillation of the $\chi$ field in
Fourier-space regions where the equation of motion is unstable. The character of the instability is
revealed by converting eq.~(\ref{eompre}) into the standard Mathieu equation. Rescaling the scalar field,
\begin{equation}
X_k = a^{3/2}\chi_k,
\end{equation}
eliminates the friction-like term and so yields
\begin{equation}
\label{eomrecast}
\ddot X_k + \bigg[ \frac{k^2}{a^2} +g^2 \Phi^2(t)\sin^2(mt) \bigg] X_k=0.
\end{equation}
Now, recasting the argument of the oscillating term by setting $ z = mt$ completes the
conversion into the Mathieu equation:
\begin{equation}
\label{matheq}
X_k'' + \left[ A_k + 2q\cos(2z) \right] X_k=0.
\end{equation}
The prime represents the derivative with respect to the argument $z$, and the two parameters in the
equation are
\begin{equation}
A_k = \frac{k^2}{a^2m^2} + 2q, \quad q = \frac{g^2\Phi^2(t)}{4m^2}.
\end{equation}
The resonance behavior of solutions to the Mathieu equation depends on the values of these $A_k$ and $q$,
which determine the stable and unstable regions. Appendix~\ref{Appendix} reproduces the standard plot
depicting the stability and instability regions in the $q$-$A_{k}$ plane with a graph of the Mathieu
equation parameters.

The oscillations of the scalar field exhibit adiabatic instability when
\begin{equation}
\label{omegainsta}
\frac{\dot \omega}{\omega^2} \gtrsim 1,
\end{equation}
and energy transfer occurs between the inflaton and the scalar field $\chi$. Trial solutions of the
Mathieu equation,
\begin{equation}\label{Xsol}
X_k \propto e^{\mu_k z},
\end{equation}
are unstable for real values of the Floquet characteristic exponent $\mu_k$~\cite{mclach1947, as1972}.
Section~\ref{Number Density} discusses $\mu_k$ in more detail.

\begin{figure}[ht!]
    \centering
    \subfloat[\centering]{{\includegraphics[width=7cm]{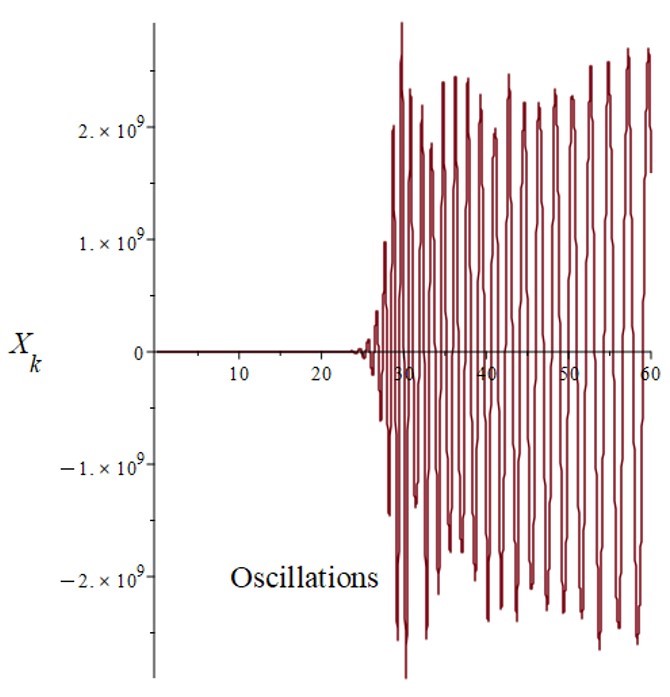} }}%
    \qquad
    \subfloat[\centering]{{\includegraphics[width=7cm]{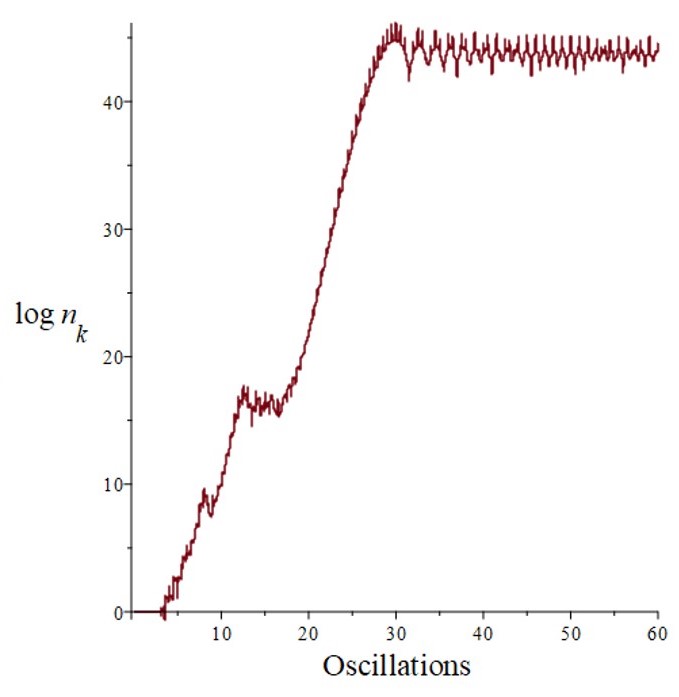} }}%
		\caption{The scalar field and occupation number for the first 60 oscillations, in the model with a
		discontinuous scale factor, and for the inflaton mass $m = 10^{-6}$ $M_{P}$. The $t$-axis is in units
		of the number of oscillations, $2\pi/m$. We have selected the specific mode of the KLS model
		with the wave number $k = 4m$ to maximize the growth of the occupation number. To reproduce the
		broad-resonance exponential growth, we have used parameters $g = 6.25 \times 10^{-4}$,
		$\dot X(t_f) = 0.045$, and $\ddot X(t_f) \approx 0$; these were identified empirically, and
		varying the parameter values away from these will decrease the observable resonance effect. The
		scalar field derivative $\dot X(t_f)$ approximates what KLS
		advise---namely, that the positive-frequency solution
		$X_k(t) \approx\exp\left(-i\omega_k t/\sqrt{2\omega_k}\right)$ be applied as an initial condition.}
    \label{Xk ln nk}
\end{figure}

The mode occupation number $n_k$ is the energy of the mode in question,
divided by the single-particle energy $\omega_k$:
\begin{equation}\label{nk}
n_k = \frac{\omega_k}{2} \left( \frac{|\dot X_k|^2}{\omega_k^2} + |X_k|^2 \right) - n_{k\,0}.
\end{equation}
(The adjustment $-n_{k\,0}$ to account for the zero-point energy density is effectively negligible.)
Figure~\ref{Xk ln nk} reproduces the results of the discontinuous scale factor of the KLS
model, for the scalar field mode amplitude $X_k$ and the exponential increase in the
corresponding occupation number $\log n_k$. The $t$-axis timeline of both graphs becomes a count of
the number of oscillations of the inflaton after $t$ is expressed in units of $\frac{2\pi}{m}$, with
which the revised equation of motion~(\ref{eomrecast}) is
\begin{equation}
\label{eomrecast2}
\ddot X_k + (2\pi)^2\left[ \frac{k^2}{a^2} +g^2 \Phi^2(t)\sin^2(2\pi t) \right] X_k=0.
\end{equation}

Broad parametric resonance preheating requires certain preconditions on $\Phi(t)$ and the Mathieu
equation parameter $q$, and it begins shortly after the end of inflation, after  approximately
one quarter of an oscillation of the inflaton. (KLS use this approximation to advance their analysis.)
With time defined in terms of the number of oscillation cycles, $t_f = \frac{\pi}{2m} \approx 10^{-37}$~s,
which makes the timeline consistent with that which we have found for the continuous scale factor; our
order-of-magnitude increase in the size of the cosmos also appears at around $10^{-37}$~s.

In graph~(b) of figure~\ref{Xk ln nk}, the scalar field spans many instability bands in the first
$\sim 10$ oscillations, as $q$ decreases substantially, and the resonances cause exponential growth in
the occupation number. From about 12 to 17 oscillations, the growth flattens as $q$ lessens while crossing
the stability region corresponding to $q$ values decreasing from about 2 to 1. Broad resonance and growth resume
in the next 10 oscillations in the instability band for $q \lesssim 1$ and $A_k \approx 1$, before ultimately
terminating after $\sim 34$ oscillations. Appendix \ref{Appendix} also shows graph~(b) of
figure~\ref{Xk ln nk} superimposed on the final three instability regions of
figure~\ref{EOM over mclachlan3} (corresponding to decreasing $q$ as time progresses).

\begin{figure}[ht!]
    \centering
    \subfloat[\centering]{{\includegraphics[width=7cm]{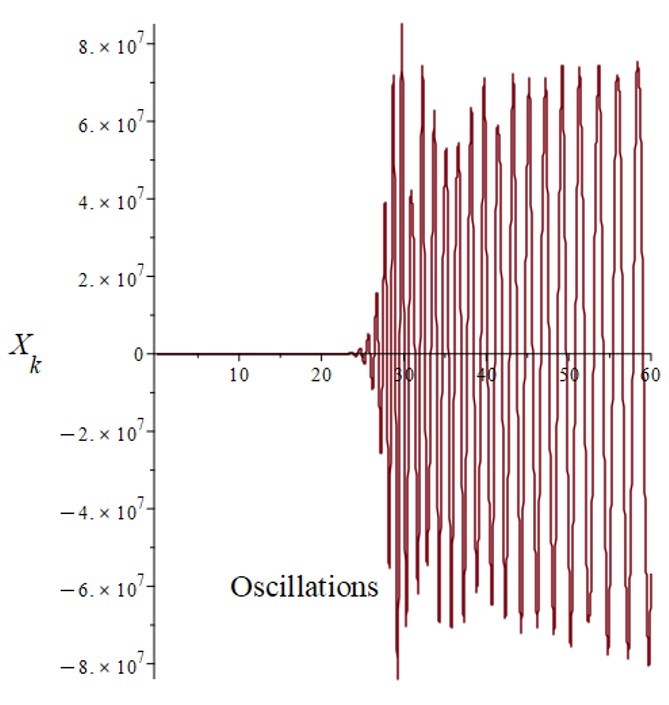} }}%
    \qquad
    \subfloat[\centering]{{\includegraphics[width=7cm]{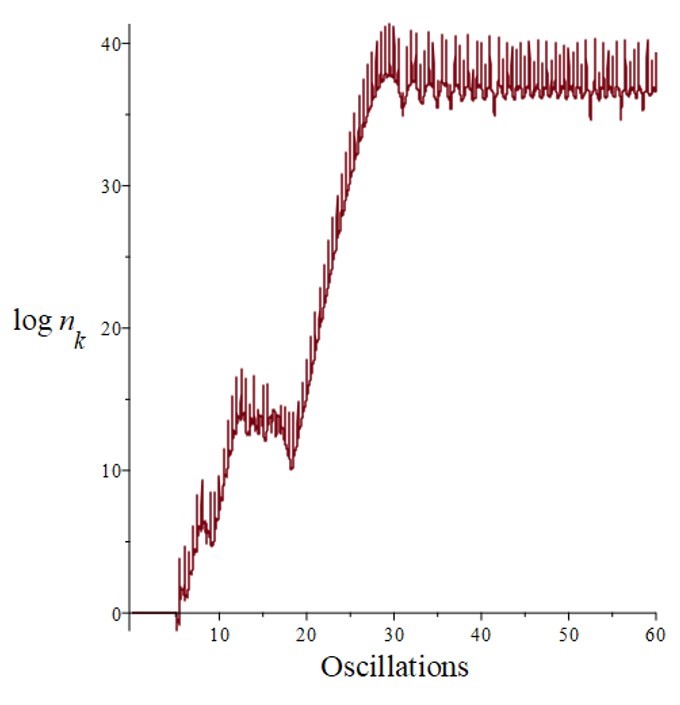} }}%
		\caption{The scalar field and occupation number for the first 60 oscillations in a
		model with the continuous scale factor. The graphs show reduced values of $X_k$ and
		$\log n_k$ compared to figure \ref{Xk ln nk} because of the effect of the order-of-magnitude
		increase in $a(t)$.}
		\label{Xk ln nk ooma}
\end{figure}

\begin{figure}[ht!]
\centering
\hspace*{-1.5cm}
\includegraphics[scale=.4950]{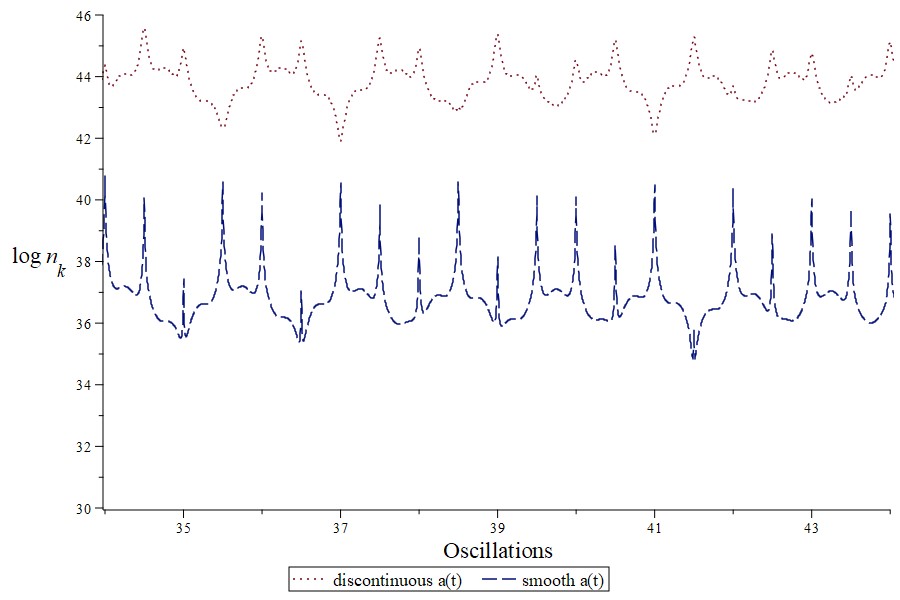}
\caption{These plots compare 10 oscillations of $\log n_k$ after the end of broad resonance at $\sim34$
oscillations for the two functional forms of the scale factor.}
\label{10oscies}
\end{figure}

In figure~\ref{Xk ln nk ooma}, we repeat the presentation from figure~\ref{Xk ln nk} using the
smooth transitional scale factor in place of the kinked scale factor of the KLS model. The scalar
field and occupation number show sharp decreases from figure~\ref{Xk ln nk} to figure~\ref{Xk ln nk ooma}.
We can examine the effect of the continuous scale factor more precisely by analyzing the root mean squares
of $X_k$ and $n_k$ averaged over the 10 oscillations following the end of preheating, which occurs
after approximately 34 oscillations. With time $t$ in units of $\text{m}^{-1}$ according to
the KLS formalism, we can convert the scale factor units of time in seconds to oscillations,
\begin{equation}
a(t) \approx \left[ \frac{t \, \text{(s)}}{t_f}\right]^{\!2/3}
=\left[\frac{t\,(\text{s})\left(\frac{2\pi/m}{\text{s}}\right)}{\frac{\pi}{2m}}\right]^{\!2/3} = (4t)^{2/3}.
\end{equation}
We have also used the assumption for $t_f$ that broad parametric preheating begins after inflation ends, at
one fourth of an oscillation. Then we apply a factor of 10 for the approximate order-of-magnitude increase
in the continuous $a(t)$,
\begin{equation}
a(t) \rightarrow  10a(t) \approx 10(4t)^{2/3}.
\end{equation}
For $X_k$ (in the $k=4m$ mode), we find a modest decline of $\sim 0.03$ in the root mean square, due to the
order-of-magnitude increase in the scale factor. Figure~\ref{10oscies} shows $\log n_k$ for both forms of
the scale factor for 10 oscillations following the end of broad resonance. The decrease in $\log n_k$
because of the effect of the larger scale factor causes a reduction of just $\sim 0.002$ in the root mean
square of the occupation number $n_k$ at 10 oscillations after broad resonance terminates.

Local maxima in $\log n_k$ for the smooth scale factor in figure~\ref{10oscies} occur at every
half oscillation of $\phi$ at $t = \frac{1}{2}, 1, \frac{3}{2},\ldots$. At these points,
where $\text{sin}(2 \pi t) = 0$ in eq.~(\ref{eomrecast2}), the frequency reduces to
\begin{equation}
\omega_k = 2\pi \frac{k}{a},
\end{equation}
with values less than one, $0.080 \lesssim \omega_k \lesssim 0.095$. For the 10 oscillation periods under
consideration with the smooth scale factor model, this range of fractional frequencies has the
effect of increasing the contribution of the term containing the kinetic energy
$\dot X_k$ in the occupation number,
\begin{equation}
n_k = \frac{\omega_k}{2} \left( \frac{|\dot X_k|^2}{\omega_k^2} + |X_k|^2 \right) - n_{k\,0},
\end{equation}
even as it tends to suppress the contribution of the potential-like $X_k$ term. Thus, the small
fractional frequency generates the local maxima. The range of larger frequencies with the cusped
scale factor following the end of resonance,
$0.80 \lesssim \omega_k \lesssim 0.95$, has less of an effect and intersperses some local minima,
depending on the relative values of $\dot{X}_k$ and $X_k$ at the half-oscillation times.

We are able to provide some understanding of the differences in appearance of $\log n_k$---that is,
the greater degree of dispersion of the amplitudes above the average occupation number in graph~(b) of
figure~\ref{Xk ln nk ooma} in comparison with figure~\ref{Xk ln nk}---by examining in detail the effect
of the fractional frequency. At oscillation 36, for example, the occupation numbers $\log n_k (36)$ are
approximately 45.3 and 40.2 for the cusped and smooth scale factor models, respectively.
The kinetic term in the energy, amplified by the frequency, for the most part determines
the occupation number in both models. The average occupation numbers over 4 oscillations from oscillation
34 to 38 are  approximately 43.9 and 36.8, respectively---yielding an increase during this period
of $\sim 0.03$ with the cusped
scale factor and $\sim 0.09$ with the smooth model. The lower level of the scalar field in the smooth
model and (more importantly) its time derivative moderate what would otherwise be an approximately
$10\text{-fold}$ difference in the increases based on the values of $\omega_k$ alone. Thus, we see the greater
dispersion of amplitudes above the average $\log n_k$ in figure \ref{Xk ln nk ooma}.  Appendix~\ref{analysis}
contains a table that lists some of the supporting data associated with the behavior around
oscillation 36, as well as related graphs.

\subsection{Number Density}
\label{Number Density}

The number density of the scalar field quanta has its basis in the process of broad parametric resonance
KLS characterize in their paper as \textit{stochastic}---that is, random. They show that the variation in
the phase $\theta_k$ of the scalar field $\chi$ in the course of semiclassical interactions between the
$\chi$-particles and the oscillating inflaton field is very much greater than $\pi$, which makes
successive phases effectively random. However, this does not mean that the there is no net energy flow
from one sector to the other. In fact, a growth in the number of particles between classical scattering
events can be as much as
3 times as probable as a decrease, based on the numerical effect of possible values for the phase angle in
the recurrence relation governing resonance. KLS also separate preheating into two time periods. The first
period precedes all backreaction and rescattering, and the second period involves the effect of those
interactions on number density, which can be significant. Backreaction and rescattering are quantum effects
in which the created $\chi$-particles interact with the background inflaton field. In backreaction,
interactions can alter the effective masses of the particles and the frequency of the inflaton
oscillations. Rescattering involves a created particle scattering again, either off an inflaton or another
$\chi$-particle. However, KLS conclude that the duration of the second period is so brief that during it they
can safely neglect the expansion of the universe, and their analysis of that part does not depend on
the scale factor. Therefore, here we shall determine the effect of the continuous scale factor on number
density conversely without including backreaction and rescattering.

Semiclassical scattering leading to quantum-mechanical $\chi$-particle production involves the
interaction of the scalar field $\chi$ and the background inflaton field oscillating around zero. KLS
derive the number density of the $\chi$ field from the adiabatic approximation solution to
eq.~(\ref{eomrecast}),
\begin{equation}
X_k(t) = \frac{\alpha_k(t)}{\sqrt{\omega_k}}e^{-i\mathlarger{\int}^{t_j} dt \, \omega_k}
+ \frac{\beta_k(t)}{\sqrt{\omega_k}}e^{+i\mathlarger{\int}^{t_j} dt \, \omega_k},
\end{equation}
with the scalar field phase $\theta_k^j = \int^{t_j} dt \, \omega_k$ and $t_j$ representing the time at
end of the $j^{\text{th}}$ oscillation---such that as time $t \rightarrow t_j$, the inflaton field
is oscillating around its minimum, $\phi \rightarrow 0$. The functions $\alpha_k(t)$ and $\beta_k(t)$
are time-dependent Bogoliubov transformation coefficients~\cite{birrell1982}.

Around $\phi \approx 0$, eq.~(\ref{eomrecast}) becomes 
\begin{equation}\label{eomphi0}
\ddot X_k + \left[ \frac{k^2}{a^2} +g^2 \Phi^2(t) m^2 (t - t_j)^2 \right] X_k=0.
\end{equation}
The scalar field $\chi$ with an effectively random phase $\theta_k^j$ completes a half-oscillation at time
$t \rightarrow t_j$ for $j = 1, 2, 3, \ldots$. As $t \rightarrow t_j$ for each half-oscillation of $\chi$,
the inflaton field  concurrently oscillates near zero, creating a period of non-adiabatic energy transfer,
which leads to exponential growth in the number of $\chi$-quanta according to eq.~(\ref{omegainsta}).
At other times, the number density $n_\chi$ remains stable. Introduction of parameters
\begin{equation}
\tau = k_*(t - t_j) \; \text{and} \; \kappa = \frac{k}{a k_*}
\end{equation}
recasts eq.~(\ref{eomphi0}) as a differential equation with a parabolic cylinder function solution,
\begin{equation}
\frac{d^{2}X_k}{d\tau^2} + (\kappa^2 + \tau^2)X_k = 0,
\end{equation}
which is also the Schr\"{o}dinger equation with an unstable quadratic potential, $V(\phi) \propto -\tau^2$.
Appendix~\ref{moderange} derives the largest mode to participate in the broad parametric resonance,
$k_* = \sqrt{gm\Phi}$. The scattering of solutions $X_k$ of eq.~(\ref{eomrecast}) leads to a
recurrence relation for the Bogoliubov coefficients, which may be represented by transfer matrix,
\begin{equation}
\begin{pmatrix} 
   \alpha_k^{j+1}e^{-i\theta_k^j}  \\
   \beta_k^{j+1}e^{+i\theta_k^j} \\
   \end{pmatrix}  = \begin{pmatrix} 
   \frac{1}{D_k} & \frac{R_k^*}{D_k^*}  \\
   \frac{R_k}{D_k} & \frac{1}{D_k^*}  \\
    \end{pmatrix} \\
	\begin{pmatrix} 
   \alpha_k^j e^{-i\theta_k^j}  \\
   \beta_k^j e^{+i\theta_k^j}  \\
   \end{pmatrix}.	
\end{equation}
KLS provide the reflection $R_k$ and transmission $D_k$ amplitudes from the solutions of the parabolic
cylinder equation and also the phase angle $\varphi_k$, which is a complicated function of the parameter
$\kappa$, 
\begin{equation}
\varphi_k = \operatorname{arg}\Gamma \left( \frac{1 + i\kappa^2}{2} \right)
+\frac{\kappa^2}{2} \left( 1 + \log \frac{2}{\kappa^2} \right).
\end{equation}
With these, the recurrence relation becomes
\begin{equation}
\begin{pmatrix} 
   \alpha_k^{j+1}  \\
   \beta_k^{j+1} \\
   \end{pmatrix}  = \begin{pmatrix} 
   \sqrt{1+e^{-\pi\kappa^2}}e^{i\pi\varphi_k} & ie^{-i\frac{\pi}{2}\kappa^2+2i\theta_k^j}  \\
   -ie^{-i\frac{\pi}{2}\kappa^2-2i\theta_k^j} & \sqrt{1+e^{-\pi\kappa^2}}e^{-i\pi\varphi_k}  \\
    \end{pmatrix} \\
	\begin{pmatrix} 
   \alpha_k^j  \\
   \beta_k^j  \\
   \end{pmatrix}.
\end{equation}
Noting that the occupation number $n_k$ in eq.~(\ref{nk}) just depends on the
Bogoliubov coefficient $\beta_k$~\cite{parker1968},
\begin{equation}
n_k = |\beta_k|^2,
\end{equation} 
and that for for a coherent process $n_k \gg 1$, leads to the recurrence relation 
\begin{equation}\label{nkrr1}
n_k^{j+1} \approx \left[ 1+2e^{-\pi\kappa^2} -2\sin(\theta_{\text{tot}}^j)\,
e^{-\frac{\pi}{2}\kappa^2} \sqrt{1 + e^{-\pi\kappa^2}}\right] n_k^j,
\end{equation} 
with the accumulated phase
\begin{equation}
\theta_{\text{tot}}^j = 2\theta_k^j - \varphi_k +\operatorname{arg}\beta_k^j -\operatorname{arg}\alpha_k^j. 
\end{equation}
Because the variation in the phases $\theta_k^j$ is very much greater than $\pi$, the randomness of
$\theta_k^j$---and by extension the randomness of $\alpha_k^j$ and $\beta_k^j$ as functions of
$\theta_k^j$---make $\theta_{\text{tot}}^j$ stochastic. Noting that resonance begins to be suppressed
unless $\pi \kappa^2 \lesssim 1$, KLS find that for $\pi \kappa^2 \ll 1$, a growth in the number of particles
is three times as likely as a decrease. Within the range $0 < \theta_{\text{tot}}^j \le 2\pi$, values of
$0 < \theta_{\text{tot}}^j < \frac{\pi}{4}$ and $\frac{3\pi}{4} < \theta_{\text{tot}}^j \le 2\pi$
cause an increase in the number of particles according to eq.~(\ref{nkrr1}); only over one quarter of
the possible range of phases, $\frac{\pi}{4} < \theta_{\text{tot}}^j \le \frac{3\pi}{4}$, does the
number of $\chi$-particles decrease, as energy flows (incoherently) back to the inflaton field.
A second recurrence relation also
obtainable~\cite{mclach1947} from the Mathieu equation~(\ref{matheq}),
\begin{equation}
n_k^{j+1} = n_k^j e^{2\pi\mu_k^j}, 
\end{equation}
in combination with eq.~(\ref{nkrr1}), yields the Floquet characteristic exponent,
\begin{equation}\label{muk}
\mu_k^j = \frac{1}{2\pi} \log \left[ 1+2e^{-\pi\kappa^2} -2 \sin(\theta_{\text{tot}}^j) \,
e^{-\frac{\pi}{2}\kappa^2} \sqrt{1 + e^{-\pi\kappa^2}}\right].
\end{equation}

Integration of $n_k$ for all modes that participate in broad parametric resonance gives rise to the
total number density of $\chi$-quanta,
\begin{equation}\label{int}
n_\chi = \frac{1}{(2\pi a)^3} \int d^3 k \, n_k(t) = \frac{1}{4\pi^2 a^3} \int dk \, k^2 e^{2\mu_k mt}.
\end{equation} 
The units of number density $n_\chi$ are the expected m$^{-3}$, since occupation number $n_k(t)$ is
dimensionless. KLS evaluate the integral on the far-right-hand side of
eq.~(\ref{int}) by the steepest descent method and
estimate the number density to be
\begin{equation}
n_\chi \approx \frac{k_*^3}{64\pi^2 a^3 \sqrt{\pi\mu mt}}e^{2\mu mt}.
\end{equation}
They also determine the maximum Floquet characteristic exponent
$\mu$ associated with an unknown maximum $k_{\text{max}}$, estimated as
$k_{\text{max}} \approx \frac{k_*}{2}$. 

We use the proportionality
\begin{equation}
\label{nxprop}
n_\chi \propto \frac{1}{a^3 \sqrt{\mu mt}}e^{2\mu mt}
\end{equation}
to perform a numerical analysis of the effect of the continuous scale factor by examining the ratio
\begin{equation}
R_{\chi}=\frac{n_{\chi \, a(t)_{\text{smooth}}}}{n_{\chi \, a(t)_{\text{cusp}}}}.
\end{equation}
The terms $n_{\chi \, a(t)_{\text{smooth}}}$ and $n_{\chi \, a(t)_{\text{cusp}}}$ represent the number
densities of the smooth and cusped scale factor models, respectively. We anticipate a decrease in the
number density due to the increase in volume, moderated to a certain amount by the dependence of the
proportionality in eq.~(\ref{nxprop}) on $\mu$. The
use of the proportionality eliminates the dependence on the unknown mode
$k_*$, which KLS estimate as $k_*(t) \approx \sqrt{gm\Phi(t)}$, as detailed in Appendix~\ref{moderange}.

\begin{figure}[ht!]
\centering
\includegraphics[scale=.55]{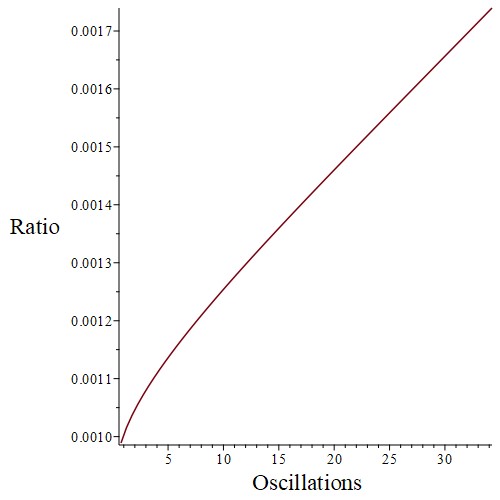}
\caption{The ratio $R_{\chi}$ of $n_\chi$ for the smooth scale factor, to that with $a(t)$ with a discontinuous
derivative at the end of inflation, as used by KLS. The increase in volume in the smooth model, resulting
from the extra time given for space to expand as inflation tails off, dilutes $n_\chi$ by $\sim 10^{-3}$,
which the broad parametric resonance term $e^{2\mu mt}/\sqrt{\mu mt}$ partially tends to offset.
Maximizing the offset with total phase $\sin(\theta_{\text{tot}}^j)=-1$ in eq.~(\ref{muk}) minimizes
the dilution, and that is what is shown in this figure. So with the smooth scale factor, $n_{\chi}$ should
be diluted by at least the ratios shown here.}
\label{ratio}
\end{figure}

In the absence of $\frac{e^{2\mu mt}}{\sqrt{\mu mt}}$, the greater time allowed for the expansion of space
in the smooth model would on its own cause dilution---that is, a decrease in the number density. The
order-of-magnitude increase in the smooth scale factor alone would reduce the number density by the cube
of the scale factor increase, $\sim 10^{-3}$. However, the effect of the broad parametric resonance in
preheating---in particular, the term $\frac{e^{2\mu mt}}{\sqrt{\mu mt}} > 1$---may modestly offset the mere
increase in the volume of space. The extent of the offset is dependent on the stochastic
$\sin(\theta_{\text{tot}}^j)$ in eq.~(\ref{muk}). Figure~\ref{ratio}
displays the ratio $R_{\chi}$ of number density of the smooth scale factor to the discontinuous scale factor
as a function of time (again expressed as the number of oscillations). The value of the ratio at the start of
preheating, $\sim 10^{-3}$, reflects the effect only of the expansion of space. As preheating progresses,
however, $R_{\chi}$ rises to a level slightly greater than $1.7 \times 10^{-3}$ at the end of broad parametric
resonance, at around 34 oscillations, in the limiting case in which $\sin(\theta_{\text{tot}}^j)$ is 
consistently equal to $-1$. In contrast, as
$\sin(\theta_{\text{tot}}^j)$ increases toward 1, $R_{\chi}$ decreases. For example, at
$\sin(\theta_{\text{tot}}^j) = 0$, $R_{\chi} \sim 1.6 \times 10^{-3}$, and $R_{\chi}$ is about
$1.3 \times 10^{-3}$ at $\sin(\theta_{\text{tot}}^j) = 0.65$. With a slightly larger stochastic
value, $n_{\chi \, a(t)_{\text{cusp}}}$ is not directly calculable via this method at lower oscillations,
and with a stochastic phase of $0.8$, the calculations of both $n_{\chi \, a(t)_{\text{cusp}}}$ and
$n_{\chi \, a(t)_{\text{smooth}}}$, even at the end of 34 oscillations, because that would
require the Floquet index $\mu$ in
eq.~(\ref{muk}) to be negative. The negative Floquet index signals an
essentially unphysical solution, which the
model formalism does not support; physically this scenario would describe a net energy flowing back into
the inflaton field, while mathematically the formalism breaks down because the saddle point integration method
is no longer usable.
Thus, examination of the ratio $R_{\chi}$ places a bound on the effect
of the continuous scale factor. The reduction of the number density due to the expansion of space alone,
$\sim 10^{-3}$, increases only slightly, by at most about $1.7 \times 10^{-3}$ after preheating,
depending on the values of the stochastic $\sin(\theta_{\text{tot}}^j)$ angles.

\section{Conclusion}

This work has explored the consequences of applying the reasonable expectation of smoothness to the
physical expansion of space, as expressed by the characteristic scale factors defining the early universe
evolving through its generally-accepted, broadly-defined epochs. We focused on the nearly instantaneous
slice of time separating the inflationary era and the subsequent era in which the stress-energy tensor
was assumed to be dominated by a single component, either radiation or matter. We focused
on the transition out of inflation specifically because it is where we inevitably expect to find the sharpest
change in the behavior of the scale factor;
assuming some realistic values for primordial parameters reveals that the time derivative of the scale
factor can decrease by a factor of $\frac{1}{120}$ between inflation and the radiation era.
Rather than being guided by a specific equation of state model,
we imposed a first-derivative smoothness requirement upon the scale factor and looked at
phenomenalistic interpolating functions that could connect the inflation and subsequent eras.
The assumption of a continuously, steadily declining (but not contracting) slope after the end of inflation
led to an in-depth examination of families of interpolating candidates with shifted power-law dependencies
on time. We imposed the same requirements of smoothness at the beginning and at the end of the brief
interpolating transition period.

From these matching conditions,
we uncovered that it was necessary to place the vertices of power law interpolating functions with indices
$n < 1$ prior to the end of inflation at $t_f$ and the vertices of functions with $n > 1$ subsequent to $t_f$,
with the displacement in either case parameterized by $\delta$.
Also initially unknown was the transition period $\Delta$---the duration of the period between the end of
inflation and the single-component universe (whether modeled as composed of radiation or matter). However,
implicit in our transition model was a remaining uncertainty in the parameters of the model. We cannot find
specific expressions for all of them without imposing additional conditions, and we can do no better than
finally expressing the displacement $\delta$ in terms of the transition period $\Delta$. Graphical analysis
of the Hubble parameter and the equation-of-state and speed-of-sound stability and causality constraints
allowed us to identify physically reasonable interpolating power-law functions as those having indices that
approached the power-law indices $\frac{1}{2}$ and $\frac{2}{3}$ for radiation and matter single-component
universes, respectively. Numerical analyses demonstrate the remarkable result that the actual transition
lasts approximately $10^{-37} \, \text{s}$, essentially regardless of the composition
of the single-component universe that follows the
transition and the duration $\Delta$. In addition, the universe enters the single-component era
about an order of magnitude (or approximately 2--3 $e$-folds) larger than it would have been if subject to
a scale factor with a discontinuous slope, which switched instantaneously to $t^{1/2}$ or $t^{2/3}$ behavior
at the end of the inflationary epoch.
Although the form of the interpolating function is not exponential, the increase in the lifespan of the
universe, $10^{-37} \, \text{s}$, is not inconsequential compared to the assumption for the inflationary expansion of the
universe, $N \approx 60$ $e$-folds. We understand the outcome to be a universe given an additional short
sliver of time in which to grow larger simply because we have imposed a condition of smoothness on the
physical expansion of space. The numerical analysis adds precision to this result. For a radiation-dominated
era following the transition, at $10^{-37} \, \text{s}$ the increase in the size of the universe has attained
$98.5\%$ of its asymptotic value, and the corresponding figure for a subsequent matter-dominated era is
$98.3\%$.
Generalizing the approach we have used to a multiple-component universe would also be interesting,
as would considering the high-scale physics of inflation that might provide the friction needed to end
inflation in a smoother way.

We proceeded to examine the effect of the theoretical changes we had described to the dynamic expansion of
space (characterized by a smooth scale factor and the resulting predicted increase in the size of the
universe) on a subsequent preheating era.
The evolution of the universe after inflation remains highly speculative because of the challenges
implicit in experimental confirmation. A period of reheating appears to be required in order to be consistent
with the later stages of cosmological development, but the details of the reheating dynamics can
depend sensitively on the nature of the particle species available to be excited---including as-yet
unobserved high-mass species that may not be accessible at standard model scales but could
nonetheless have been active participants in the dynamics of the hot, dense early universe.
However, we have also discussed the intricate, highly-technical theory of preheating developed by
Kofman, Linde, and Starobinsky to address some generic problems with reheating.
We applied the KLS formalism to our model with a smooth interpolating scale factor leading into
a matter-dominated universe, in order to gauge
the effects of the smoothing on the most sensitive $\chi$-particle occupation number $n_{k}$
and the corresponding number density $n_\chi$. We were able to estimate the numerical changes compared with
the results obtained using the standard cusped
scale factor, and we concluded that the differences are not necessarily numerically significant, apart
from a dilution in the total particle density that should be common to all models that predict somewhat
larger universes after the end of inflation. Specifically, for the
occupation number of the most aggressively growing $\chi$ mode, we find a modest decline in $\log n_{k\,=\,4m}$
of $\sim 2 \times 10^{-3}$ in the root mean square for 10 oscillations following the end of broad parametric
resonance, which is a consequence of a decrease of just $\sim 3 \times 10^{-2}$ in the root mean square of
the scalar field $\chi_k$ over the same period. In addition, by constructing a relation consisting of the
ratio of the number density in the cosmology with the smooth scale factor to that with the cusped
scale factor, we determine a partial offset to the expected dilution of the quantity of bosons produced
by broad parametric resonance due to the approximate $10^3$ increase in the unit volume of space caused
by the larger smooth scale factor. The stochastic nature of broad parametric resonance precludes a
specific prediction, but we find an additional modest increase in the proportion,
with an upper bound of $\lesssim 1.7 \times 10^{-3}$. 

It may be somewhat
surprising that the effect of a proposed smoothing of the scale factor is so minor---mostly
limited to the natural rarefaction of the $\chi$ particles that comes with a spatially larger universe.
Regarding the possibility (in a case of optimal phase alignment) of, at most, an additional near doubling
per unit volume of number density, we note that a doubling of a small number of something in a unit volume
may easily be thought of as not negligible. However, in terms of the many, many orders of magnitude of
primal particles in a unit volume of early space, we consider the outcome of having, at most, close to
twice as many as not of substance. Thus, we view the result of the numerical analysis of the effect of a not
insignificant increase in the size of the universe to represent confirmation of the comparative
invariance of the KLS preheating model to these kinds of modifications.
We are satisfied that result should represent a modestly useful
contribution to the body of work in support of this iconic theory.

\begin{appendices}

\section{}
\label{temp}

In this appendix, we review the derivation of
the inverse relation between the radiation temperature of the universe and the
scale factor, following the approach outlined by Ryden~\cite{ryden2003}. In an isothermal environment, the
first law of thermodynamics
\begin{equation}
dQ = dE + p \, dV
\end{equation}
reduces to 
\begin{equation}
\label{firstlaw}
\frac{dE}{dt} = -p\frac{dV}{dt}.
\end{equation}
After substituting the pressure of a relativistic gas, $p_\gamma = \frac{\rho_\gamma}{3}$,
and the CMB black body energy density, $\rho_\gamma =4\sigma T^{4}$, we get
\begin{align}
\frac{d (\rho_\gamma V)}{dt} &= -\frac{\rho_\gamma}{3} \frac{dV}{dt}, \\
\frac{d \rho_\gamma}{dt} V + \rho_\gamma\frac{d V}{dt} &= -\frac{\rho_\gamma}{3} \frac{dV}{dt}, \\
\frac{1}{T}\frac{dT}{dt} &= -\frac{1}{3V}\frac{dV}{dt}.
\end{align}
In an expanding universe with volume element $V = a^3(t)L^3$, the relation becomes
\begin{equation}
\frac{1}{T}\frac{dT}{dt} = -\frac{1}{3a^3(t)L^3}\frac{d[a^3(t)L^3]}{dt}=-\frac{1}{a}\frac{da}{dt},
\end{equation}
which is an elementary separable differential equation, satisfied for
\begin{equation}
T \propto a^{-1}.
\end{equation}

\section{}
\label{eHworking}

This appendix details the derivation of the working forms of the equation of state in flat space,
defined as
\begin{equation}
\label{eHapp1}
\epsilon_H = -\frac{\dot{H}}{H^2} = -\frac{\ddot{a}/a - (\dot{a}/a)^2}{(\dot{a}/a)^2}
= 1 - \frac{\ddot{a}/a}{(\dot{a}/a)^2}.
\end{equation}
Substituting the first Friedmann equation in flat space,
\begin{equation}
\label{Feq1}
H^2(t) = \left( \frac{\dot{a}}{a} \right)^{\!\!2} = \frac{8\pi G}{3}\rho,
\end{equation}
and the acceleration equation,
\begin{equation}
\frac{\ddot{a}}{a} =  -\frac{4\pi G}{3}(\rho + 3p),
\end{equation}
gives
\begin{equation}
\epsilon_H = 1 - \frac{\frac{-4\pi G}{3} (\rho + 3p)}{\frac{8\pi G}{3} \rho} = \frac{3}{2} (1 + \omega),
\end{equation}
where $\omega$ would be the coefficient of proportionality between pressure $p$ and density $\rho$ in
a single-component universe. For an exponential scale factor $a = e^{Ht}$, we clearly have
$d\log a = H dt$, but in fact this relationship holds more generally, as integrating it just gives the
definition of $H$. So we can alternatively express $\epsilon_{H}$ as
\begin{equation}
\epsilon_H = -\frac{\dot{H}}{H^2} = \frac{1}{H} \left( -\frac{\dot{H}}{H} \right)
= \frac{1}{H}  \frac{d}{dt} \log (H^{-1}) = \frac{d \log \left(H^{-1}\right)}{d \log a},
\end{equation}
that is, as the slope along a plot of $\log \left(H^{-1}\right)=-\log H$ versus $\log a$.

\section{}
\label{singcomp}

The derivation of $\epsilon_{H\,\text{expected}} = \frac{1}{n}$ for a single-component universe is:
\begin{align}
a_n(t) &= a_n(t_0) \left( \frac{t}{t_0} \right)^{\!\!n} \\
\dot{a}_n(t) &= n \frac{a_n(t_0)}{t_0^{\, n}} t^{(n - 1)} = \frac{n}{t} a_n(t) \\
\ddot{a}_n(t) &= n \frac{a_n(t_0)}{t_0^{\, n}} (n - 1) t^{(n - 2)} = \frac{n - 1}{t} \dot{a}_n(t).
\end{align}
Therefore,
\begin{align}
\epsilon_H &= -\frac{\dot{H}}{H^2} = 1 - \frac{\ddot{a}/a}{(\dot{a}/a)^2}
= 1 - \frac{a_n(t) \ddot{a}_n(t)}{\dot{a}_n(t)^2} \\
&= 1 - \frac{a_n(t) \frac{n - 1}{t}\dot{a}_n(t)}{\dot{a}_n(t)^2}
= 1 - \frac{a_n(t) \frac{n - 1}{t}}{\frac{n}{t} a_n(t)} = \frac{1}{n}.
\end{align}


\section{}
\label{Appendix}

We reproduce here the well-known stability-instability
chart~\cite{mclach1947,as1972} showing the regions of the parameter space in which the initial
value problem solutions for the Mathieu equation~(\ref{matheq}) are either stable or unstable. 

\begin{figure}[ht!]
\hspace*{2.8cm}
\includegraphics[scale=0.75]{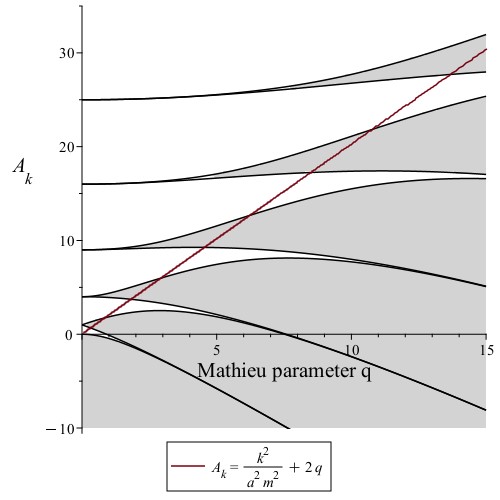}
\caption{Stability-instability chart. The areas highlighted in gray are the regions of instability
in the $q$-$A_{k}$ parameter space. The plot also depicts the Mathieu equation parameters associated with
the equation of motion solutions. Note that $q$ and $A_{k}$ decrease as time progresses.}
\label{EOM over mclachlan3}
\end{figure}

\begin{figure}[ht!]
\hspace*{.8cm}
\includegraphics[scale=0.60]{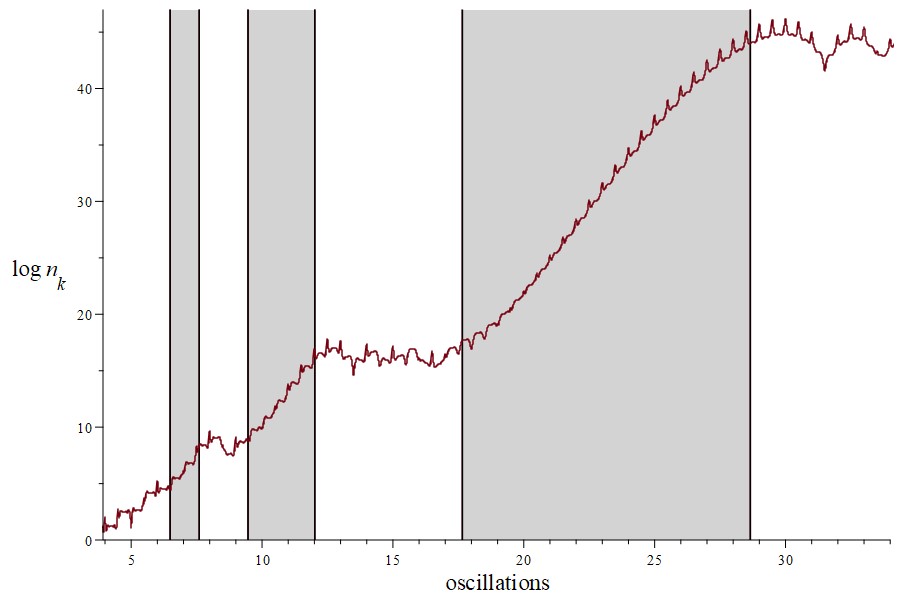}
\caption{The final three instability regions superimposed on the $\log n_k$ resonance growth. As the number
of oscillations increases, we see exponential growth in occupation number as the $q $ and $A_k$ of
figure~\ref{EOM over mclachlan3} decline toward zero and the equation of motion crosses the last three
instability regions.}
\label{overlay q}
\end{figure}

The regions of instability correspond to the periods of sustained exponential growth in $n_{k}$
during preheating.
Figure~\ref{overlay q}
superimposes the Mathieu equation instability regions associated with the equation of motion
in the KLS model on the time evolution of $\log n_k$, to highlight this correspondence.

There are some interesting and noteworthy differences between the behavior in the three broad-resonance
growth regions for $n_{k}$ (at least for the particular, rapidly-growing value $k=4m$ we have selected).
There are small oscillations visible, in addition to the secular growth in $\log n_k$.
During periods when the parameters make the Mathieu equation stable (the white bands in
figure~\ref{overlay q}), the oscillations are comparatively chaotic; this is also what was seen in
figure~\ref{10oscies} after the last resonant growth period has ending. There is a certain amount of
approximately periodic behavior, due to the driving by the amplitude squared of the inflaton field, so
there are fairly stark features every half an inflaton oscillation period.  However, underneath these is
a chaotically varying baseline.  During the periods of resonance (the gray bands), the baseline behavior
is different, with approximately exponential growth in the occupation number, as is typical in an
unstable driven system. On top of this are additional oscillations, qualitatively similar in some ways
to those in the stability regions. However, there are also clear manifestations of the nonlinearity of the
Mathieu equation, in the form of period doubling or tripling. When the exponential growth is
subtracted, the residual still has, on average, one peak per half
oscillation of the inflaton field. However, these peaks are not evenly placed or of equal amplitude.
During the second shown resonance region, the oscillating residuals have periods equal to
the full inflaton oscillation period---a period doubling phenomenon. Within each full oscillation are two
dissimilar up-and-down cycles. Moreover, in the vicinity of and during the first, shortest resonant period
there is period tripling, with the periodic residuals taking one and half inflaton oscillation cycles to
return fully to their original phase space positions.

\section{}\label{analysis}

\begin{table}[H]\centering  
\begin{tabular}{l@{}ccccccccccc@{}}\toprule  
& $a(t) \, \text{model}$ & \phantom{a} & $\omega_k$ & \phantom{a}& $\omega_k |X|^2$ & \phantom{a}
& $ \frac{|\dot X|^2}{\omega_k}$ & \phantom{a} & $\log n_k (36)$ & \phantom{a} & Increase \\ \midrule
& cusped & \phantom{a} & $0.91$ & \phantom{a} & $4.62 \times 10^{17}$ & \phantom{a} & $9.56 \times 10^{19}$
& \phantom{a} & $45.3$ & \phantom{a} & $0.03$\\ [.15cm] 
 & smooth & \phantom{a} & $0.091$ & \phantom{a} & $8.71 \times 10^{13}$ & \phantom{a} & $5.79 \times 10^{17}$
& \phantom{a} & $40.2$& \phantom{a} & $0.09$\\ 
\bottomrule  
\end{tabular}
\caption*{Table 5: Data in support of the differences in appearance between graphs~(b) of
figures~\ref{Xk ln nk} and~\ref{Xk ln nk ooma}. The last column represents the increase in occupation
number $\log n_k (36)$ compared to the average value over 4 oscillations from oscillation
34 to 38, which are $43.9$ and $36.8$ for the KLS and smooth scale factor models, respectively.}
\label{disp}    
\end{table}

\begin{figure}[ht!]
\centering
\hspace*{-1.5cm}
\includegraphics[scale=.4950]{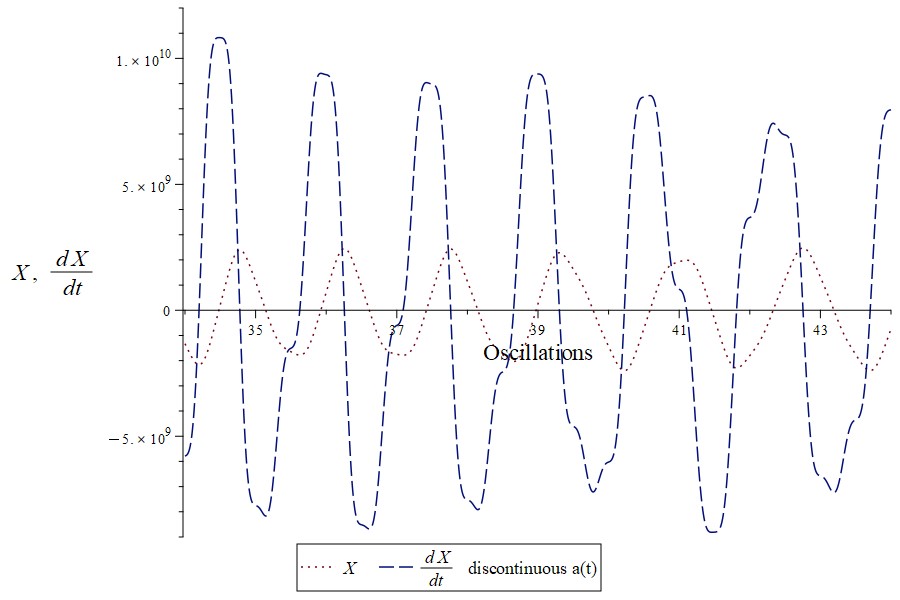}
\caption{Plots of scalar field $X$ and its time derivative $\dot X$ with the cusped scale factor for
10 oscillations following the end of broad parametric resonance.}
\label{XDX dis}
\end{figure}

\begin{figure}[ht!]
\centering
\hspace*{-1.5cm}
\includegraphics[scale=.4950]{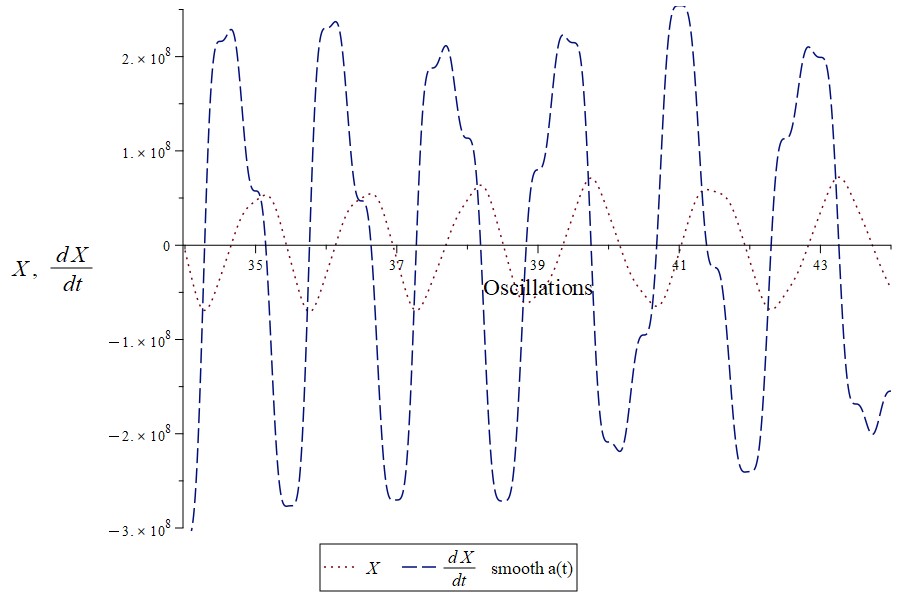}
\caption{Plots of scalar field $X$ and its time derivative $\dot X$ with the smooth scale factor for
10 oscillations following the end of broad parametric resonance.}
\label{XDX con}
\end{figure}

\newpage

\section{}
\label{moderange}

\begin{figure}[ht!]
\hspace*{2.5cm}
\includegraphics[scale=0.65]{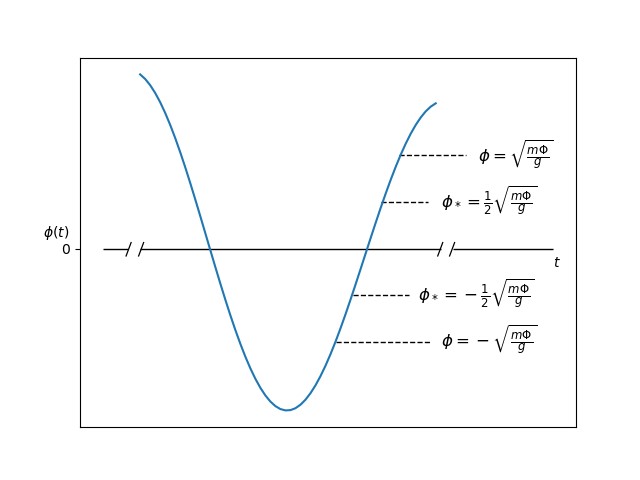}
\caption{The outer and inner pairs of dotted lines represent the  ranges of $\phi(t)$ that participate
in parametric resonance. The wider outer band corresponds to the values of $\phi(t)$ that
participate in the resonance for the
minimal Fourier component---that is, as $k \rightarrow 0$. The inner band,
$-\frac{1}{2}\sqrt{m\Phi/g} \le \phi_* \le \frac{1}{2}\sqrt{m\Phi/g}$, corresponds to 
the participating $\phi(t)$ associated for the modes with $k_* = \sqrt{gm\Phi}$. This is the preheating
band of broad parametric resonance. Explosive growth in the number of particles occurs as
$\phi(t) \rightarrow 0$.}
\label{bands}
\end{figure}

Appendix \ref{moderange} details the derivation of the range of modes
$k_*$ that participate in the broad parametric resonance process. KLS provide the typical frequency for the
scalar field $\chi$ oscillations, $\omega(t) = \sqrt{k^2 + g^2\phi^2(t)}$, subject to the adiabatic
instability condition, eq.~(\ref{omegainsta}),
\begin{equation}
\frac{\dot \omega}{\omega^2} \gtrsim 1
\end{equation}
\begin{equation}
\frac{\dot \omega}{\omega^2} = \frac{g^2\phi\dot\phi}{\left( k^2 + g^2\phi^2 \right)^{3/2}} \gtrsim 1.
\end{equation}
The instability condition yields the inequality defining the unstable modes. The inflaton at the
end of inflation is an oscillating field of the form $\phi(t) = \Phi(t)\sin(mt)$. For broad resonance,
when $\phi(t)$ is small and the decaying envelope $\Psi$ is approximately constant over the period of
a single oscillation, $\dot{\phi} \approx m\Phi$. This makes the resonance condition
\begin{align}
\label{inequal1}
1 & \lesssim \frac{g^2\phi m\Phi}{\left( k^2 + g^2\phi^2 \right)^{3/2}} \\
\label{inequal2}
k^2 & \lesssim \left( g^2\phi m\Phi \right)^{2/3} - g^2 \phi^2.
\end{align}
We find the maximum range of $k$ by taking the derivative of the inequality (\ref{inequal2}) to
maximize the inflaton value $\phi_*$,
\begin{equation}
\frac{2g^{4/3}m^{2/3}\Phi^{2/3}}{3\phi_*^{1/3}} - 2g^2 \phi_* = 0,
\end{equation}
for which the solution is
\begin{equation}
\phi_* =(3)^{-3/4}\sqrt{\frac{m\Phi}{g}} \approx \frac{1}{2}\sqrt{\frac{m\Phi}{g}}.
\end{equation}
Substituting $\phi_* \approx \frac{1}{2}\sqrt{\frac{m\Phi}{g}}$ into
$k_{\text{max}}^2 \lesssim \left( g^2\phi_* m\Phi \right)^{2/3} - g^2 \phi_*^2$, we find
\begin{equation}
k_{\text{max}} \lesssim \sqrt{\frac{gm\Phi}{2}}.
\end{equation}
Taking $k \rightarrow 0$ in eq.~(\ref{inequal1}) generates an expression for the
inflaton associated with the minimum-range of mode $k$:
\begin{equation}
\frac{g^2\phi m\Phi}{\left( g^2\phi^2 \right)^{3/2}} \gtrsim 1
\end{equation}
\begin{equation}
\phi \lesssim \sqrt{\frac{m\Phi}{g}}.
\end{equation}

Figure~\ref{bands} shows a standard graphical representation of the bands of $\phi(t)$ associated with
the  minimum and maximum ranges of $k$. We note that $k_{\text{max}}^2$ applies to a band of $\phi(t)$
for which $|\phi| \le 2\phi_*$. Thus, we find
\begin{equation}
k_* = \sqrt{gm\Phi}.
\end{equation}

\end{appendices}


\begin{thebibliography}{99}

\raggedright
\bibitem{ryden2003}
B. Ryden, \textit{Introduction to Cosmology} (Addison Wesley, San Francisco, 2003).
\bibitem{LiddleLyth2006}
A. R. Liddle and D. H. Lyth, \textit{Cosmological Inflation and Large-Scale Structure}
(Cambridge University Press, New York, 2006).
\bibitem{linde1984}
A. D. Linde, ``The Inflationary Universe,'' Rep. Prog. Phys. \textbf{47}, 925 (1984).
\bibitem{baumann2012}
D. Baumann, ``TASI Lectures on Inflation,'' {\tt arXiv:0907.5424}.
\bibitem{lesgourgues2006}
J. Lesgourgues, ``Inflationary Cosmology,'' Lecture Notes, {\tt https://lesgourg.github.io/courses.html}.
\bibitem{liddle2015}
A. R. Liddle, \textit{An Introduction to Modern Cosmology} (Wiley \& Sons, West Sussex, 2015).
\bibitem{bassett2006}
B. A. Bassett, S. Tsujikawa, and D. Wands, ``Inflation Dynamics and Reheating,'' Rev. Mod. Phys.
\textbf{78}, 537 (2006); {\tt arXiv:astro-ph/0507632}.
\bibitem{kinney2004}
W. H. Kinney, ``Cosmology, Inflation, and the Physics of Nothing,'' {\tt arXiv:astro-ph/0301448}.
\bibitem{weinberg2008}
S. Weinberg, \textit{Cosmology} (Oxford University Press, Oxford, 2008).
\bibitem{fixsen2009}
D. J. Fixsen, ``The Temperature of the Cosmic Microwave Background,'' Astrophys. J. \textbf{707},
916 (2009); {\tt arXiv:0911.1955}.
\bibitem{PlIV2018}
N. Aghanim, \textit{et al.}, ``Planck 2018 results. VI. Cosmological parameters,'' Astron. Astrophys.
\textbf{641}, A6 (2020); {\tt arXiv:1807.06209}.
\bibitem{friedmann1922}
A. Friedmann, ``About the Curvature of Space,'' J. Phys. \textbf{10}, 377 (1922).
\bibitem{hooft1974}
G. ’t Hooft, ``The Inflationary Universe,'' Nucl. Phys. B \textbf{79}, 276 (1974).
\bibitem{polyakov1974}
A. M. Polyakov, ``Particle Spectrum in Quantum Field Theory,'' JETP Lett. \textbf{20}, 194 (1974).
\bibitem{guth1981}
A. H. Guth, ``The Inflationary Universe: A Possible Solution to the Horizon and Flatness
Problems,'' Phys. Rev. D \textbf{23}, 347 (1981).
\bibitem{linde1982}
A. D. Linde, ``A New Inflationary Universe Scenario: A Possible Solution of the Horizon,
Flatness, Homogeneity, Isotropy and Primordial Monopole Problems,'' Phys. Lett. B \textbf{108},
389 (1982).
\bibitem{linde2007}
A. D. Linde, ``Inflationary Cosmology,'' Lect. Notes Phys. \textbf{738}, 1 (2008);
{\tt arXiv:0705.0164}.
\bibitem{linde1983}
A. D. Linde, ``Chaotic Inflation,'' Phys. Lett. B \textbf{129}, 177 (1983).
\bibitem{lyth2007}
D. H. Lyth, ``Particle physics models of inflation,'' Lect. Notes Phys. \textbf{738}, 81 (2008);
{\tt arXiv:hep-th/0702128}.
\bibitem{riotto2017}
A. Riotto, ``Inflation and the Theory of Cosmological Perturbations,'' {\tt arXiv:hep-ph/0210162}.
\bibitem{kls1997}
L. Kofman, A. D. Linde, and A. A. Starobinsky, ``Towards the Theory of Reheating after
Inflation,'' Phys. Rev. D {\bf 56}, 3258 (1997); {\tt arXiv:hep-ph/9704452}.
\bibitem{kolb1999}
E. W. Kolb, ``Dynamics of the Inflationary Era,'' {\tt arXiv:hep-ph/9910311}.
\bibitem{allahverdi2010}
R. Allahverdi, R. Brandenberger, F. Y. Cyr-Racine, and A. Mazumdar, ``Reheating in Inflationary Cosmology:
Theory and Applications,'' Ann. Rev. Nucl. Part. Sci. \textbf{60}, 27--51 (2010); {\tt arXiv:1001.2600}.
\bibitem{linde2002}
A. D. Linde, ``Inflationary Cosmology and Creation of Matter in the Universe,'' in S. Bonometto, V. Gorini,
and U. Moschella (eds.), \textit{Modern Cosmology} (Institute of Physics Publishing, Bristol, UK, 2002),
p. 159--185.
\bibitem{lozanov2019}
K. Lozanov, ``Lectures on Reheating after Inflation,'' {\tt arXiv:1907.04402}.
\bibitem{weinberg2009}
S. Weinberg, ``Effective Field Theory, Past and Future,'' Proc. Sci. \textbf{CD09}, 001 (2009);
{\tt arXiv:0908.1964}.
\bibitem{lindblom2018}
L. Lindblom, ``Causal Representations of Neutron-Star Equations of State,'' Phys. Rev. D \textbf{97},
123019 (2018); {\tt arXiv:1804.04072}.
\bibitem{ellis2007}
G. F. R. Ellis, R. Maartens, and M. A. H. MacCallum,``Causality and the Speed of Sound,'' Gen. Rel. Grav.
\textbf{39}, 1651 (2007); {\tt gr-qc/0703121}.
\bibitem{thorne2017}
K. S. Thorne and R. D. Blandford, \textit{Modern Classical Physics} (Princeton University Press,
Princeton, 2017).
\bibitem{mclach1947}
N. W. McLachlan, \textit{Theory and Applications of Mathieu Functions} (Oxford University Press, London,
1947).
\bibitem{as1972}
M. Abramowitz and I. Stegun, \textit{Handbook of Mathematical Functions} (Dover, New York, 1972).
\bibitem{birrell1982}
N. D. Birrell and P. C. W. Davies, \textit{Quantum Fields in Curved Space} (Cambridge University Press,
Cambridge, 1982).
\bibitem{parker1968}
L. Parker, ``The Creation of Particles by the Expanding Universe,'' Ph.D. thesis, Harvard University (1966).

\end{thebibliography}
\end{document}